\documentclass{ametsocV6.1}

\makeatletter
\newcommand{\myitem}[1]{%
\item[#1]\protected@edef\@currentlabel{#1}%
}
\makeatother
\title{Capturing the Variability of the Nocturnal Boundary Layer through Localized Perturbation Modeling}

\authors{Amandine Kaiser,\aff{a}\correspondingauthor{Amandine Kaiser, amandine.kaiser@geo.uio.no}
Nikki Vercauteren,\aff{a,b}
Sebastian Krumscheid,\aff{c}}

\affiliation{\aff{a}{Department of Geosciences, University of Oslo, Oslo, Norway}\\
\aff{b}{Institut für Geophysik und Meteorologie, Universität zu Köln, Cologne, Germany}\\
\aff{c}{Department of Mathematics, Karlsruhe Institute of Technology, Karlsruhe, Germany}}

\abstract{A single-column model is used to investigate regime transitions within the stable atmospheric boundary layer, focusing on the role of small-scale fluctuations in wind and temperature dynamics and of turbulence intermittency as triggers for these transitions. Previous studies revealed abrupt near-surface temperature inversion transitions within a limited wind speed range. However, representing these transitions in numerical weather prediction (NWP) and climate models is a known difficulty. To shed light on boundary layer processes that explain these abrupt transitions, the Ekman layer height and its correlation with regime shifts are analyzed. A sensitivity study is performed with several types of perturbations of the wind and temperature tendencies, as well as with the inclusion of intermittent turbulent mixing through a stochastic stability equation. The effect of small fluctuations of the dynamics on regime transitions is thereby quantified. The combined results for all tested perturbation types indicate that small-scale phenomena can drive persistent regime transitions from very to weakly stable regimes, but for the opposite direction, no evidence of persistent regime transitions was found. The inclusion of intermittency prevents the model from getting trapped in the very stable regime, thus preventing the so-called "runaway cooling", an issue for commonly used short-tail stability functions. The findings suggest that using stochastic parameterizations of boundary layer processes, either through stochastically perturbed tendencies or parameters, is an effective approach to represent sharp transitions in the boundary layer regimes and is, therefore, a promising avenue to improve the representation of stable boundary layers in NWP and climate models.}

\begin{document}
\nolinenumbers
%% Necessary!
\maketitle

%%%%%%%%%%%%%%%%%%%%%%%%%%%%%%%%%%%%%%%%%%%%%%%%%%%%%%%%%%%%%%%%%%%%%
% SIGNIFICANCE STATEMENT/CAPSULE SUMMARY
%%%%%%%%%%%%%%%%%%%%%%%%%%%%%%%%%%%%%%%%%%%%%%%%%%%%%%%%%%%%%%%%%%%%%
%
% If you are including an optional significance statement for a journal article or a required capsule summary for BAMS 
% (see www.ametsoc.org/ams/index.cfm/publications/authors/journal-and-bams-authors/formatting-and-manuscript-components for details), 
% please apply the necessary command as shown below:
%
% Significance Statement (all journals except BAMS)
%
%\statement
%	 Enter significance statement here, no more than 120 words. See \url{www.ametsoc.org/index.cfm/ams/publications/author-information/significance-statements/} for details.
%
%% Capsule (BAMS only)
%%
%\capsule
%       Enter BAMS capsule here, no more than 30 words. See \url{www.ametsoc.org/index.cfm/ams/publications/author-information/formatting-and-manuscript-components/#capsule} for details.
%
%% * * If using twocol mode, you will need to use the commands "twocolsig" and "twocolcapsule" in place of "sig" and "capsule"
%%      to ensure that the text box correctly spans across both columns.
%

%%%%%%%%%%%%%%%%%%%%%%%%%%%%%%%%%%%%%%%%%%%%%%%%%%%%%%%%%%%%%%%%%%%%%
% MAIN BODY OF PAPER
%%%%%%%%%%%%%%%%%%%%%%%%%%%%%%%%%%%%%%%%%%%%%%%%%%%%%%%%%%%%%%%%%%%%%
%

%% In all cases, if there is only one entry of this type within
%% the higher level heading, use the star form: 
%%
% \section{Section title}
% \subsection*{subsection}
% text...
% \section{Section title}

%vs

% \section{Section title}
% \subsection{subsection one}
% text...
% \subsection{subsection two}
% \section{Section title}

%%%
% \section{First primary heading}

% \subsection{First secondary heading}

% \subsubsection{First tertiary heading}

% \paragraph{First quaternary heading}

\section{Introduction}
The stably stratified atmospheric boundary layer (SBL) is commonly classified into two stability regimes. The first one is a weakly stable regime, which corresponds to strong wind, weak stratification, and relatively continuous turbulence in both space and time, while the second one is very stable and is characterized by low wind, strong stratification, and weak or intermittent turbulence \citep{mahrt_stably_2014}. In the very stable regime, turbulence is not strong enough to mix the whole layer, which leads to the decoupling of the lower from the higher levels \citep{derbyshire_boundary-layer_1999, van_de_wiel_minimum_2012,acevedo_contrasting_2015}. As a result, the potential temperature difference between the surface and a higher level is large. In contrast, in the weakly stable regime, turbulence is sufficiently strong to keep the layers coupled, and therefore, the potential temperature difference is low \citep{vignon_stable_2017}.

A conceptual model for near-surface temperature inversions showed that through feedbacks between the surface thermal processes and turbulent fluxes, the relationship between the temperature inversion strength and wind speed can be non-monotonic and exhibit a Z-shape dependency (\cite{van_de_wiel_regime_2017}, see also figure \ref{fig:bif_diagram}b). For low wind speeds, the \cite{van_de_wiel_regime_2017} model has only one solution with a strong inversion. In contrast, for high wind speeds, the inversion strength is small. For intermediate wind speeds, when the thermal coupling between the surface and the atmosphere is weak, three inversions of strength ranging from strong to weak are possible solutions of the conceptual model, leading to the Z-shape dependency of the inversion on the wind speed. This non-linear dependency is commonly interpreted as a sign that the SBL is bistable, i.e., two coexisting stable equilibria exist for a fixed wind speed forcing, separated by an unstable equilibrium. This Z form has, for example, been observed empirically at Dome C, Antarctica, by \cite{vignon_stable_2017}, and the existence of an unstable equilibrium has been hypothesized to explain the scatter found in the observational data in a certain range of intermediate wind speeds \citep{van_de_wiel_regime_2017}. In principle, bistability should be analyzed using a fixed forcing parameter. However, the previous studies used a reference height as a so-called velocity crossing level, where the wind was empirically found to remain rather constant throughout a night \citep{van_hooijdonk_shear_2015}. To analyse the impact of this reference height on the observed nonlinearity, another metric that is independent of the choice of the height is used. The chosen metric is the Ekman layer height and the growth of this layer is analyzed under different forcing scenarios. By studying the growth of the Ekman layer, the goal is to detect inherent nonlinearity in the dynamics of the SBL. In particular, field analyses highlighted the existence of a wind threshold where turbulence intensity increases rapidly \citep{sun_turbulence_2012}, and such rapid increase could lead to a nonlinear growth of the layer. The Ekman layer height growth is hence chosen as a proxy for the SBL dynamics and its dependence of the forcing wind speed is analysed.

Transitions between the two SBL regimes are frequently observed in SBL dynamics worldwide \citep{abraham_climatological_2019}. However, the representation of regime structures and transitions in numerical weather prediction (NWP) and climate models is inadequate, primarily due to coarse resolution (both vertical and horizontal) and an imperfect understanding of the various physical processes governing the SBL \citep{holtslag_stable_2013, sandu_why_2013}. A typical solution to include turbulent mixing due to unresolved processes is to enhance the mixing \citep{sandu_why_2013}, but this solution blurs the sharp transitions \citep{baas_near-neutral_2017}. The lack of understanding of the SBL has so far prevented the emergence of a better solution. An imperfect understanding of processes leads to model uncertainties, and in fact numerous approaches have been proposed to incorporate model uncertainty in NWP. A particularly attractive solution is to use stochastic parameterization \citep{palmer_nonlinear_2001, christensen_constraining_2020}. Stochastic parameterizations recognize the importance of model uncertainty to accurately represent the mean state, and in particular they have the potential to induce regime transitions in cases where a system exhibits multiple coexisting equilibria. This makes their incorporation crucial for a statistically more faithful representation of the mean state of the system \citep{berner_stochastic_2017}. Indeed, with a stochastic parameterization, a wider range of scenarios can be explored compared to a deterministic one, and therefore, multiple co-existing equilibria can be obtained in the range of outcomes. Stochastic parameterization approaches are divided into randomized tendencies and stochastic parameterizations. The European Center for Medium-Range Weather Forecasts (ECMWF) uses, for example, a scheme that stochastically perturbs tendencies. The perturbations are added to the model a posteriori, i.e., after it has been tuned in the deterministic mode \citep{palmer_primacy_2017}. Adding the stochasticity after the tuning has the undesirable potential to detune it. Alternatively, a priori perturbation of uncertain parameters can be used. In this case the model is tuned with the stochastic scheme included. This approach is called stochastic parameterization and does not have the detuning issues arising in the sense of perturbed tendency approaches. In this work, the impact of perturbed tendencies and stochastic parameterization on regime transitions will be analyzed.

In a previous investigation conducted by \cite{kaiser_sensitivity_2024}, a modified version of the conceptual model proposed by \cite{van_de_wiel_regime_2017} was employed as a representative model to examine the sensitivity of the polar stable boundary layer (SBL) to small-scale fluctuations and explore the influence of model uncertainty on the average state of the boundary layer. The findings of the aforementioned study demonstrated that random perturbations can induce transitions in the SBL regimes and impact the timing of these transitions. The model by \cite{van_de_wiel_regime_2017} is a strongly idealized model with no spatial dimension, and hence does not allow for spatially localized perturbations. The current study extends the research by \cite{kaiser_sensitivity_2024} using a single-column model with spatial and time dimensions. The added spatial dimension allows for more degrees of freedom for the nonlinear growth of initially small perturbations, and thereby provides a greater opportunity to capture the real-world complexities and variability of the SBL. An example where spatially localized phenomena are non-linearly amplified is given by \cite{lan_turbulence_2022}. The authors showed that small-scale wind profile distortions can cause very to weakly stable regime transitions. Therefore, this study analyzes the effect of perturbed wind dynamics in the single-column model. Scenarios where the perturbations are added above or below the Ekman layer are studied separately. \cite{sun_turbulence_2012} showed that turbulence intermittency can occur when top-down turbulent events diffuse into an environment characterized by weak turbulence. Hence, top-down mixing events can be important feedback processes for recoupling the SBL.
The direct numerical simulations of \cite{donda_collapse_2015} support the addition of perturbations to the model, as they showed that transient perturbations can be sufficient to recover a turbulent regime, i.e., for transitions from very to weakly stable regimes. Intermittent turbulence events are another possible set of processes that can trigger transitions \citep{mahrt_stably_2014}. These events have been observed to play a significant role in turbulent transport, particularly in highly stable conditions \citep{acevedo_core_2003}. Hence, we study the effect of including intermittent events through a stochastic tubulence parameterization on regime transitions. This stochastic parameterization is defined by randomizing the stability correction used in Monin Obukhov similarity theory (MOST) to scale the mixing length of the turbulence with a dimensionless stability of the atmosphere.

Vertical mixing associated with surface heterogeneity, gravity waves, or mesoscale variability is commonly not represented in NWP models \citep{sandu_why_2013}. One approach to enhancing the mixing properties of the atmospheric boundary layer in NWPs is using long-tail stability functions in MOST. The issue with these types of stability functions is that they produce too much mixing compared to observations \citep{chechin_effect_2019}, resulting in a suppression of the backfolding of the before mentioned Z curve when plotting the temperature inversion over wind speed \citep{van_de_wiel_regime_2017, baas_near-neutral_2017}. On the other hand, short-tail stability functions, which suppress turbulence in the very stable case, are more consistent with turbulence theory \citep{chechin_effect_2019}, but they can cause "runaway cooling" \citep{mahrt_stratified_1998, steeneveld_current_2014, kahnert_advancing_2022}.

One suggestion on how to circumvent the issues with stability functions was made by \cite{maroneze_new_2023}. There, the authors suggest an adaption of a classic atmospheric boundary layer parameterization, which does not use a prescribed stability function. Instead, a turbulent heat flux prognostic equation ensures the stratification dependence on other characteristics of the mean and turbulent flows. While this approach is promising, it relies on additional equations in NWPs. 

As an alternative solution that does not add prognostic equations, we suggest using stochastic modeling. Indeed, transitions from very to weakly stable regimes do not seem to be determined by internal or external state variables in a predictable manner \citep{abraham_climatological_2019}. This suggests that parameterizations that facilitate such transitions in weather and climate models may need to explicitly incorporate stochastic features. Moreover, including stochasticity is one way to account for unresolved variability due to coarse resolution that could be highly relevant for triggering large-scale transitions. Therefore, in the present work, we suggest using a short-tail stability function with added stochasticity to prevent the model from getting "stuck" in the very stable regime. A particular useful example of such a randomized stability function was developed by \cite{boyko_stochastic_2023}. Their stochastic stability equation adjusts the turbulent mixing length based on the dimensionless stability used in MOST. Since this dimensionless stability is a ratio of buoyant to shear forces, it is a pragmatic way to examine and model the combined effect of buoyancy and shear perturbations.

\cite{abraham_prototype_2019} created a stochastic model featuring transition probabilities dependent on stratification, while \cite{ramsey_empirical_2022} formulated a data-driven stochastic differential equation model to capture the variability of temperature inversion.

To study transitions in the SBL and specifically analyse the impact of localized perturbations on transitions, the paper will answer the following three questions:
\begin{enumerate}
    \item When a regime transition is observed at a fixed reference height, how does the Ekman layer height evolve?
    \item What types of perturbation of the dynamics are particularly relevant triggers for regime transitions?
    \item Can stochastic modeling be used to represent the impact of small-scale fluctuations effectively in NWP and climate models?
\end{enumerate}
Question one is meant to investigate the inherent non-linearity of the SBL without relying on a fixed observation height. Questions two and three address where models and parameterizations must be improved to represent regime transitions better, which is a particularly relevant consideration for NWPs.

In Section \ref{sec:model}, the single-column model is introduced and described in detail. As part of this section, the Ekman layer height and its relation to the geostrophic wind and regime transitions is analyzed. Following that, in Section \ref{sec:sensitivity_analysis}, two approaches are used to explore the impact of small-scale perturbations on regime transitions and identify the most impactful variables. First, in section \ref{sec:sensitivity_analysis}\ref{sec:perturbed_wind_tempearture}, a sensitivity analysis of the single-column model to small-scale fluctuations in the temperature and wind dynamics for different perturbation sizes is presented. In Section \ref{sec:sensitivity_analysis}\ref{sec:perturbed_wind_tempearture} part \ref{sec:perturbation_def} the perturbations are defined, in Section \ref{sec:sensitivity_analysis}\ref{sec:perturbed_wind_tempearture} part \ref{sec:effect_perturbation_size}, the impact of the perturbation size is quantified, and in Section \ref{sec:sensitivity_analysis}\ref{sec:perturbed_wind_tempearture} part \ref{sec:effect_perturbation}, four different stability scenarios are analyzed in detail. Then, in Section \ref{sec:sensitivity_analysis}\ref{sec:ekman_layer_perturbed}, the Ekman layer height growth is studied for the perturbed model. Lastly, Section \ref{sec:sensitivity_analysis}\ref{sec:random_stab_func} studies the impact of a randomized stability function. For this, the stability function, which is used in the rest of the paper, is replaced with the stochastic one, which was defined by \cite{boyko_stochastic_2023}.  

\section{Single-column model}\label{sec:model}
\subsection{Model description}\label{sec:model_description}
\begin{figure}[t]
  \noindent\includegraphics[width=\textwidth,angle=0]{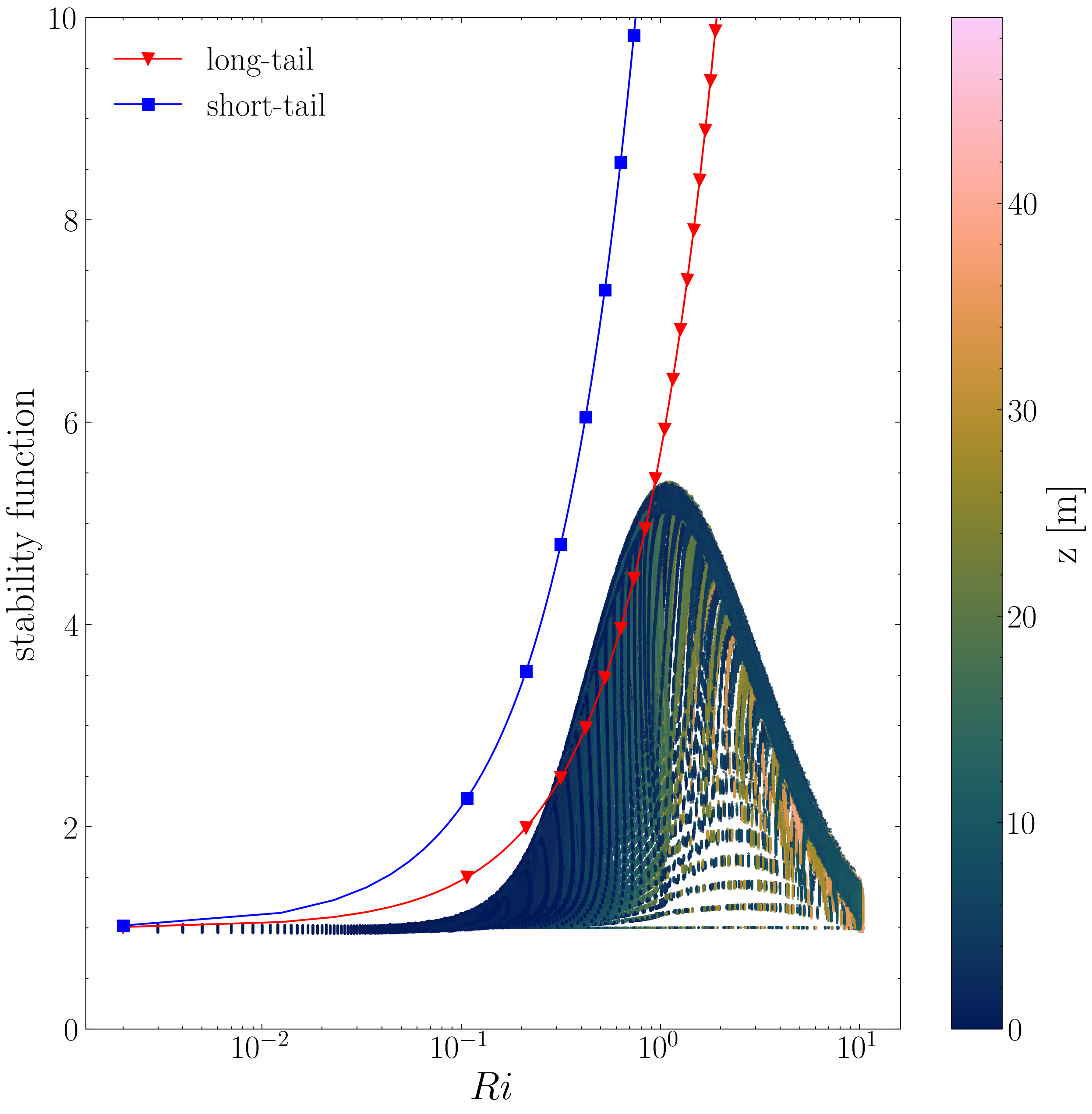}\\
  \caption{Plot of the three stability correction approaches over the Richardson number. The two lines correspond to the deterministic stability functions and the points realisations of the stochastic stability equation with a high noise strength, i.e., $\sigma_s=1.0$. The stochastic stability equation depends on the height, $z$, indicated by color. Only values for $z\leq 50$~m are displayed.}\label{fig:stab_funcs}
\end{figure}

To study which quantities can potentially trigger regime transitions in NWP and climate models, an idealized single-column model (SCM) is used, to which perturbations will be added. This model is an adapted version of a classic Reynolds-averaged Navier–Stokes 1.5-order closure model for the atmospheric boundary layer \citep{stull_introduction_1988}. It is briefly introduced below, but a more detailed description is given in \cite{boyko_simulating_2023}. The idealized SBL is defined as:
\begin{align}
		\frac{\partial u}{\partial t} &= f_c \cdot (v - v_G) + \frac{\partial}{\partial z}\left(K_m(\phi) \frac{\partial u}{\partial z}\right) -N_u\label{eq:pde_u}\\
		\frac{\partial v}{\partial t} &= f_c \cdot (u_G - u) + \frac{\partial}{\partial z}\left(K_m(\phi) \frac{\partial v}{\partial z}\right) -N_v\label{eq:pde_v}\\
		\frac{\partial \theta}{\partial t} &= \frac{\partial}{\partial z}\left(K_h(\phi) \frac{\partial \theta}{\partial z}\right) \label{eq:pde_theta}\\
		\frac{\partial k}{\partial t} &= \frac{\partial}{\partial z}\left(K_m(\phi) \frac{\partial k}{\partial z}\right)+K_m(\phi)\left(\left(\frac{\partial u}{\partial z}\right)^2+\left(\frac{\partial v}{\partial z}\right)^2\right)-\frac{g}{\theta_0}K_h(\phi)\frac{\partial \theta}{\partial z} - \frac{(\alpha_{\epsilon}k)^{3/2}}{l_m(\phi)}\label{eq:pde_tke}\\
		\frac{d \theta_g}{d t} &= \frac{1}{c_g}(R_n - H_0) - \kappa_m(\theta_g - \theta_m), \quad t>0, \quad z=0 \label{eq:pde_theta_g}
\end{align}
where $u$ and $v$ are the Reynolds averaged horizontal wind components, $\theta$ the Reynolds averaged potential temperature, $k$ the Reynolds averaged turbulent kinetic energy (TKE), and $\theta_g$ the surface temperature whose evolution is represented through a surface energy balance and modeled using a force-restore method \citep{stull_introduction_1988, garratt_review_1994, acevedo_external_2021}. The thermal capacity of the soil per unit area is given by $c_g$, $H_0 = \rho c_p \overline{w^\prime \theta_0^\prime}$ is the surface sensible heat flux, where $\rho$ is the air density, $c_p$ is the specific heat of air at constant pressure and $\theta_0$ is a reference temperature. The parameter $\kappa_m=1.18\omega$ is the soil heat transfer coefficient, with $\omega$ being the Earth’s angular frequency. The net radiation is given by $R_n$, and $\theta_m$ is the soil temperature below the surface. Additionally, $u_G$ and $v_G$ are the geostrophic wind components, and $f_c$ is the Coriolis parameter. Moreover, $g$ is the gravitational force, and $\alpha_{\epsilon}$ is a modeling constant set to 0.1.  To damp inertial oscillations, the relaxation terms 
\begin{equation*}
    N_u = \frac{u-u_G}{\tau_r}, \quad N_v = \frac{v-v_G}{\tau_r} \\
\end{equation*}
are subtracted from the momentum equations \eqref{eq:pde_u} and \eqref{eq:pde_v}. They nudge the solution towards the geostrophic wind. To ensure only a mild nudging, $\tau_r$ is set to five hours, equating to roughly 13\% of the oscillation period of the velocity components around the equilibrium wind. Lastly, the exchange coefficients for momentum $K_m$ and heat $K_h$ are defined as
\begin{equation*}
	K_m = \alpha l_m \sqrt{k}, \quad K_h = \frac{K_m}{Pr}
\end{equation*}
with a Prandtl number, $Pr$, set to 1. The mixing length is defined as $l_m = \frac{\kappa z}{\phi(Ri)+\frac{\kappa z}{\lambda}}$ where $\phi$ is a stability correction function and $\lambda=2.7\cdot 10^{-4}\frac{u_G}{|f_c|}$ \citep{rodrigo_investigation_2013} and $\kappa=0.41$ is the von Karman constant. The gradient Richardson number, $Ri$, is defined as
\begin{equation}
    Ri = \frac{\frac{g}{\theta_0}\frac{\partial \theta}{\partial z}}{\left(\frac{\partial u}{\partial z}\right)^2+\left(\frac{\partial v}{\partial z}\right)^2}. \label{eq:richardson_number}
\end{equation}
Two deterministic stability functions, $\phi$, are considered
\begin{align}
    \phi&=1+12Ri, \label{eq:short_tail_stab_func}\\
    \phi&=1+4.7Ri \label{eq:long_tail_stab_func}
\end{align}
\citep{delage_parameterising_1997}. For an increase in stability, i.e., an increase in Ri, the first stability function decreases the mixing length quicker compared to the second. Hence, we define the first function as a short-tail stability function and the second function to be a long-tail one. In addition, a third stability correction approach is tested. This stability correction is dynamically given by a data-driven stochastic stability equation initially suggested by \cite{boyko_stochastic_2023} and will be used to model random fluctuations in the turbulent mixing within the model. It is randomized in time and depends on the height, $z$. Figure \ref{fig:stab_funcs} shows the two stability functions along with the stochastic stability equation. The
stochastic stability equation is discussed in detail in section \ref{sec:sensitivity_analysis}\ref{sec:random_stab_func}.

The model equations \eqref{eq:pde_u} to \eqref{eq:pde_tke} are discretized in space using a Finite Element method and the time discretization is done with an implicit Euler scheme. For the Finite Element method, the Python package FEniCS (https://fenicsproject.org/) is used. Equation \eqref{eq:pde_theta_g} is solved using an explicit Euler algorithm. A power-three transformation with $N_z$ grid points is applied along the z-axis to enhance the resolution of all four gradients near the surface. Near the surface, the resolution is the highest with 3~cm, and close to the domain top it is the lowest with 8.5~m.

The same values as in \cite{boyko_simulating_2023} are used for all parameters, unless stated otherwise. All parameter values are summarized in Table \ref{tab:model_param} for convenience. Further details about the model and how it is numerically approximated can be found in \cite{boyko_simulating_2023}.

\begin{table}[t]
\caption{Default parameters for the model (eqs. \eqref{eq:pde_u}--\eqref{eq:pde_theta_g}).}\label{tab:model_param}
\begin{center}
\begin{tabular}{l|c|c|c}
\hline
\textbf{Description} & \textbf{Parameter} & \textbf{Value} & \textbf{Unit}\\
\hline \hline
Time step size & $\Delta t$ & 10 & s\\
Number of z grid points & $N_z$ & 100 & -\\
Roughness length & $z_0$ & 0.044 & m\\
Roughness length for heat & $z_{0h}$ & $z_0\cdot 0.1$ & m\\
Domain height & $H$ & 300 & m\\
\hline
Latitude & $l$ & 40 & $^{\circ}$\\
Coriolis parameter & $f_c$ & $2 \cdot 7.27 \cdot 1e^{-5} \cdot \sin(l\pi / 180)$ & $^{\circ}s^{-1}$\\
Geostrophic wind & $u_G$ & - & $\mathrm{ms^{-1}}$\\
Geostrophic wind & $v_G$ & 0 & $\mathrm{ms^{-1}}$\\
Relaxation time scale & $\tau_r$ & $3600 \cdot 5$ & s\\
\hline
Gravitational acceleration & $g$ & 9.81 & $\mathrm{ms^{-1}}$\\
Reference potential temperature & $\theta_0$ & 300 & K\\
Dissipation constant & $\alpha_{\epsilon}$ & 0.1 & -\\
\hline
Angular earth velocity & $\omega$ & $2 \pi / (24 \cdot 3600)$ & $^{\circ}s^{-1}$\\
Thermal capacity of soil per unit area & $c_g$ & $0.95 \cdot (1.45 \cdot 3.58 \cdot 1e^{6} / 2 / \omega) ^{0.5}$ & J$\mathrm{m^{-2}K^{-1}}$\\
Net radiation & $R_n$ & -30 & W$\mathrm{m^{-2}}$\\
Air density & $\rho$ & 1.225 & kg$\mathrm{m^{-3}}$\\
Air specific heat capacity & $c_p$ & 1005 & J$\mathrm{kg^{-1}K^{-1}}$\\
Restoring temperature & $\theta_m$ & 300 & K\\
\hline
\end{tabular}
\end{center}
\end{table}

\subsection{Initial and boundary conditions}\label{sec:initial_bound_cond}
The initial and boundary conditions are similar to the ones in \cite{boyko_simulating_2023}. They are repeated here for completeness. These initial conditions are used to initialize the non-perturbed runs. The perturbed runs (section \ref{sec:sensitivity_analysis}) are initialized with the quasi-stationary state, which is defined in section \ref{sec:model}\ref{sec:steady_state}, of the non-perturbed runs.

The initial condition for $u$ is a logarithmic profile $u(z, t_0)= \frac{u^*}{\kappa}\ln(\frac{z}{z_0})$ where $u^*_{init}=\sqrt{0.5c_f u_G^2}$ is the initial friction velocity and $c_f=4 \cdot 10^{-3}$ is a tuning parameter to ensure the wind velocity is close to $u_G$ at the upper boundary. The other wind velocity component, $v$, is constant and equal to zero as an initial condition. The initial condition for $\theta$ is given by
\begin{equation}
    \theta(z, t_0) = \begin{cases}
        \theta_0 &\text{for } z\leq z_c\\
        \Gamma (z-z_c)  + \theta_0 & \text{otherwise,}
    \end{cases}
\end{equation}
where $z_c=200$~m and $\Gamma=0.01$ is the atmospheric lapse rate. Following \cite{parente_comprehensive_2011} the initial profile for the TKE, $k$, is defined as 
\begin{equation}
    k(z, t_0)=a\ln(z+z_0)+b.
\end{equation}
The coefficients $a$ and $b$ are estimated using the boundary values:
\begin{align}
    k(z_0, t_0)&=\frac{{u^*_{z_0}}^2}{\sqrt{0.087}},\\
    k(H, t_0)&=0,\\
    u^*_{z_0}&=\frac{\kappa}{\ln(z_1/z_0)}\sqrt{u(z_1)^2+v(z_1)^2},
\end{align}
where $H$ is the domain height and $z_1$ is the next grid point after $z_0$ which is the roughness length. The initial condition for the surface temperature, $\theta_g$, is equal to $\theta_0$ and for the wind speed, v, it is equal to 0.

The boundary conditions at the lower boundary, i.e., $z=z_0$, are Dirichlet boundary conditions and are given by
\begin{align}
    u(z_0,t)&=0,\\
    v(z_0,t)&=0,\\
    \theta(z_0,t)&=\theta_g(t)\;\text{, and}\\
    k(z_0,t)&= k(z_0, t_0).
\end{align}
The upper boundary condition, $z=H$, for $v$ is zero. For all other variables, the upper boundary condition is a Neumann condition. The gradient of $\theta$ equals the lapse rate, and all other gradients are set to zero.

\subsection{Quasi-stationary state analysis}\label{sec:steady_state}
\begin{figure}[t]
  \noindent\includegraphics[width=\textwidth,angle=0]{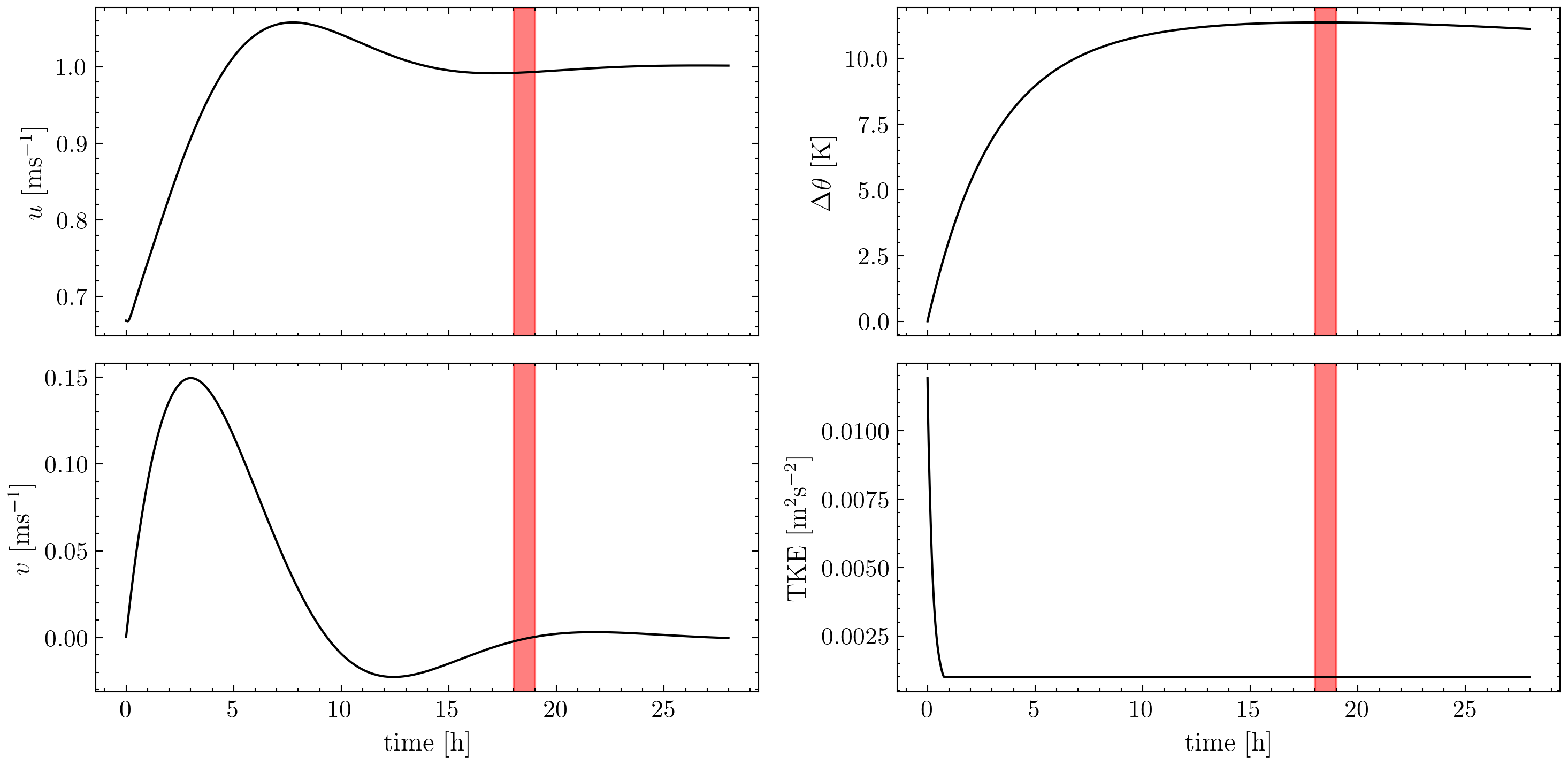}\\
  \caption{Plots of $u, \ \Delta \theta, \ v$ and TKE at $z_{ss}=20$ m for $u_G=1.0 \ \mathrm{ms^{-1}}$. The quasi-stationary state period, which is the quasi-stationary state plus one hour, is shaded in red.}\label{fig:steady_state_ex}
\end{figure}

The system is run into a quasi-stationary state before analyzing the impact of transient phenomena. This state is assumed to be reached when the solutions of equations \eqref{eq:pde_u} -- \eqref{eq:pde_tke} are nearly constant from this state on for at least one hour. To be precise, to calculate the quasi-stationary state for a variable $X$, i.e., $u, v, \Delta \theta$ or $k$, at a fixed height $z_{ss}$, a rolling 5-hour mean of $X$ is calculated. This is denoted as $\overline{X_{t_{ss}-5h}}$. A quasi-stationary state is reached at the time point $t_{ss}$ when 
\begin{equation*}
    |\overline{X_{{t_{ss}}-5h}} - X_{t_{ss}}| \leq 0.05.
\end{equation*}
This means at the quasi-stationary state, the variable deviates at most by 0.05 to the average of the previous 5 hours for a fixed height, $z_{ss}$, (see as an example figure \ref{fig:steady_state_ex} a)). The first ten simulation hours are considered the initializing period and are therefore discarded for this analysis. The length of the rolling mean window is chosen purely for modeling purposes. For simplicity, the maximal difference between the rolling mean of the variables is set to a fixed value. It is the smallest value for which, for all investigated $u_G$, a quasi-stationary state is found within a 90~h simulation. As the quasi-stationary state is only determined to initialize the perturbed runs with the values of all four model variables at this state, the thresholds to compute the quasi-stationary state are of minimal importance as long as the system no longer changes significantly at this state. 

In the following discussion, the geostrophic wind speed, which is defined as $s_G=\sqrt{u_G^2+v_G^2}$, is used instead of $u_G$. It shall be noted, though, that $s_G$ is equal to $u_G$ as $v_G$ is set to zero for all simulations. As we compare the inversion strength for a range of geostrophic wind forcings in the following section, we choose one fixed height, $z_{ss}$, for all $s_G$ to determine the quasi-stationary state. This height is the same at which the effect of perturbations on regime transitions is studied in Section \ref{sec:sensitivity_analysis}, i.e. $z_{ss}=20$~m. A fixed height is chosen to make the results comparable to the ones of other studies (e.g.,  \cite{abraham_climatological_2019_1}). To analyze the sensitivity to perturbations, we use the model's values at the quasi-stationary state as initial conditions for the perturbed runs.

\subsection{Stability regimes}\label{sec:bifurcation_analysis}
\begin{figure}[t]
  \noindent\includegraphics[width=\textwidth,angle=0]{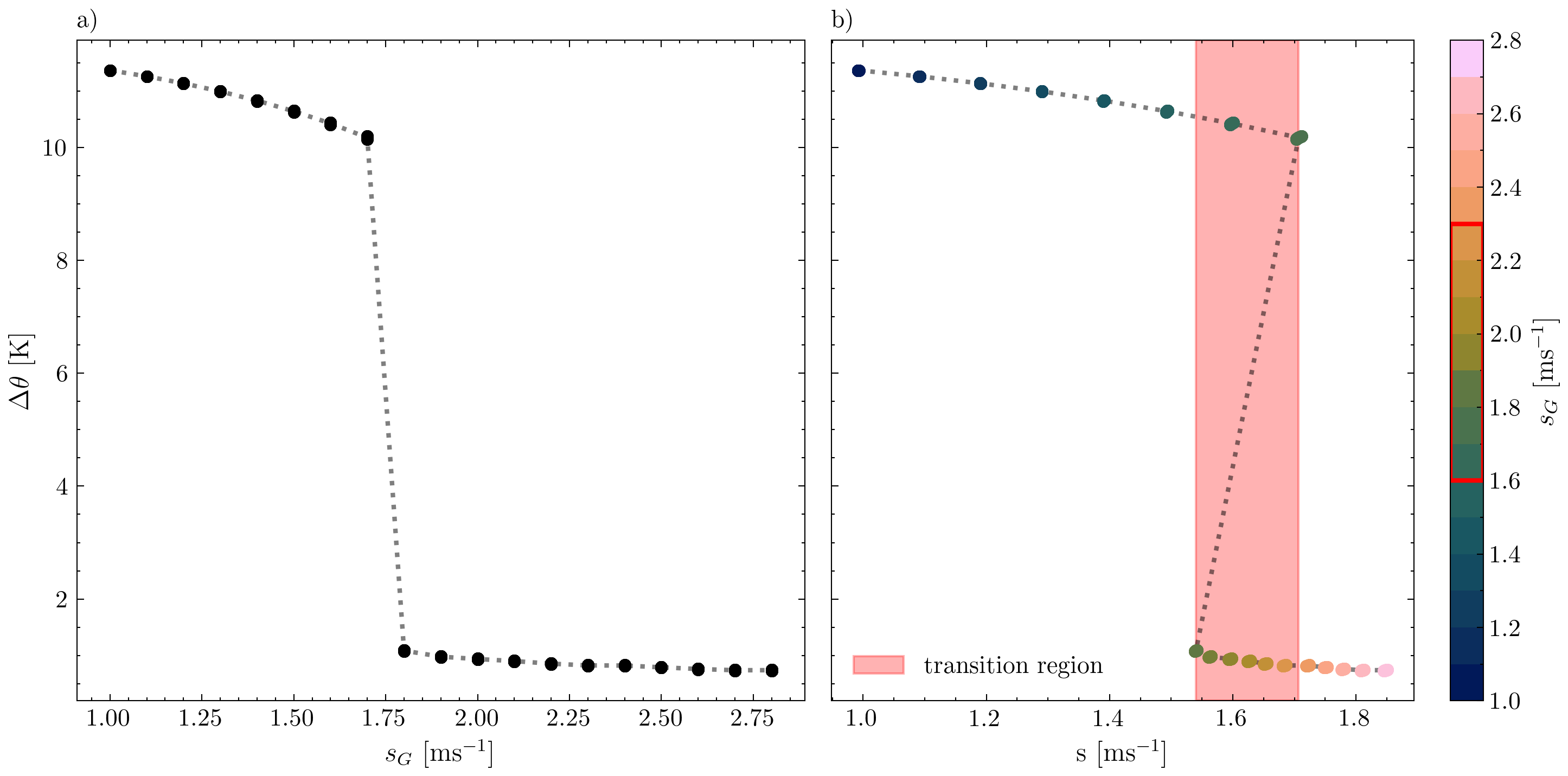}\\
  \caption{Plots of the quasi-stationary state of $\Delta \theta$ over $s_G$ (a) and $s$ at a height of 20 meters (b). In plot b), the different colors show the different forcing values, $s_G$. The dashed lines mark the mean $\Delta \theta$ value for each $s_G$ or $s$. The transition region, highlighted in red, is accompanied by a red box marking the corresponding $s_G$ values in the colorbar.}\label{fig:bif_diagram}
\end{figure}

In the literature, when the plot of the temperature inversion over wind speed resembles a Z-shape with a region where the same wind speed corresponds to two different temperature differences (as seen in figure \ref{fig:bif_diagram}b), this pattern is called "backfolding". It is commonly interpreted as a sign of bistability of the SBL\citep{van_de_wiel_regime_2017, vignon_stable_2017}. This implies that there are forcing scenarios for which the system's resilience to perturbations is weaker, which might facilitate transitions after the perturbation in these cases. \cite{van_de_wiel_regime_2017} showed that the strength of thermal coupling with the surface determines if the SBL can exhibit bistability. A weak thermal coupling, such as occurring over thermally insulating snow cover, leads to a bistable SBL in their conceptual study.

To study if the backfolding can be observed for the model used in this study (eq. \eqref{eq:pde_u}-\eqref{eq:pde_theta_g}), the model is run for a range of geostrophic wind forcings. When plotting the quasi-stationary state of $\Delta \theta_{z_{ss}}$ over the wind speed $s_{z_{ss}}$, which is defined as $s_{z_{ss}}=\sqrt{u_{z_{ss}}^2+v_{z_{ss}}^2}$, for all geostrophic wind speed forcings, $s_G$, (see figure \ref{fig:bif_diagram}b) the graph shows the backfolding when a reference height $z_{ss}=20$~m is chosen. Following the arguments made by \cite{van_de_wiel_regime_2017}, the backfolding in figure \ref{fig:bif_diagram}b indicates that the solution of model \eqref{eq:pde_u}--\eqref{eq:pde_theta_g} is bistable with the before mentioned parameters. If the model is bistable, it should have two stable equilibrium solutions and one unstable one for specific wind speeds. Based on this plot, low wind speeds and high-temperature differences characterize the first stable regime; the other one has high wind speeds and low-temperature differences. The first regime is, therefore, very stable, and the second is weakly stable. The potential bistable region, i.e., the region where for the same $u_{z_{ss}}$ both stable states are a possible solution, is shaded in red in figure \ref{fig:bif_diagram}b. The transition wind speed, i.e., the geostrophic wind speed that separates the two regimes, is 1.7 $\mathrm{ms^{-1}}$. This agrees with results from \cite{acevedo_external_2021} where the authors showed that for FLOSS II data, the wind speed associated with regime transitions can be as low as 1.0 $\mathrm{ms^{-1}}$ for a net radiation of roughly -30 $\mathrm{Wm^{-2}}$, which is the net radiation used in this study.

If not specified differently, $\Delta \theta$ and $u$ refer to $\Delta \theta_{ss}$ and $u_{ss}$ in the following discussion. The $ss$ index is dropped for notational convenience. 

\begin{figure}[t]
  \noindent\includegraphics[width=\textwidth,angle=0]{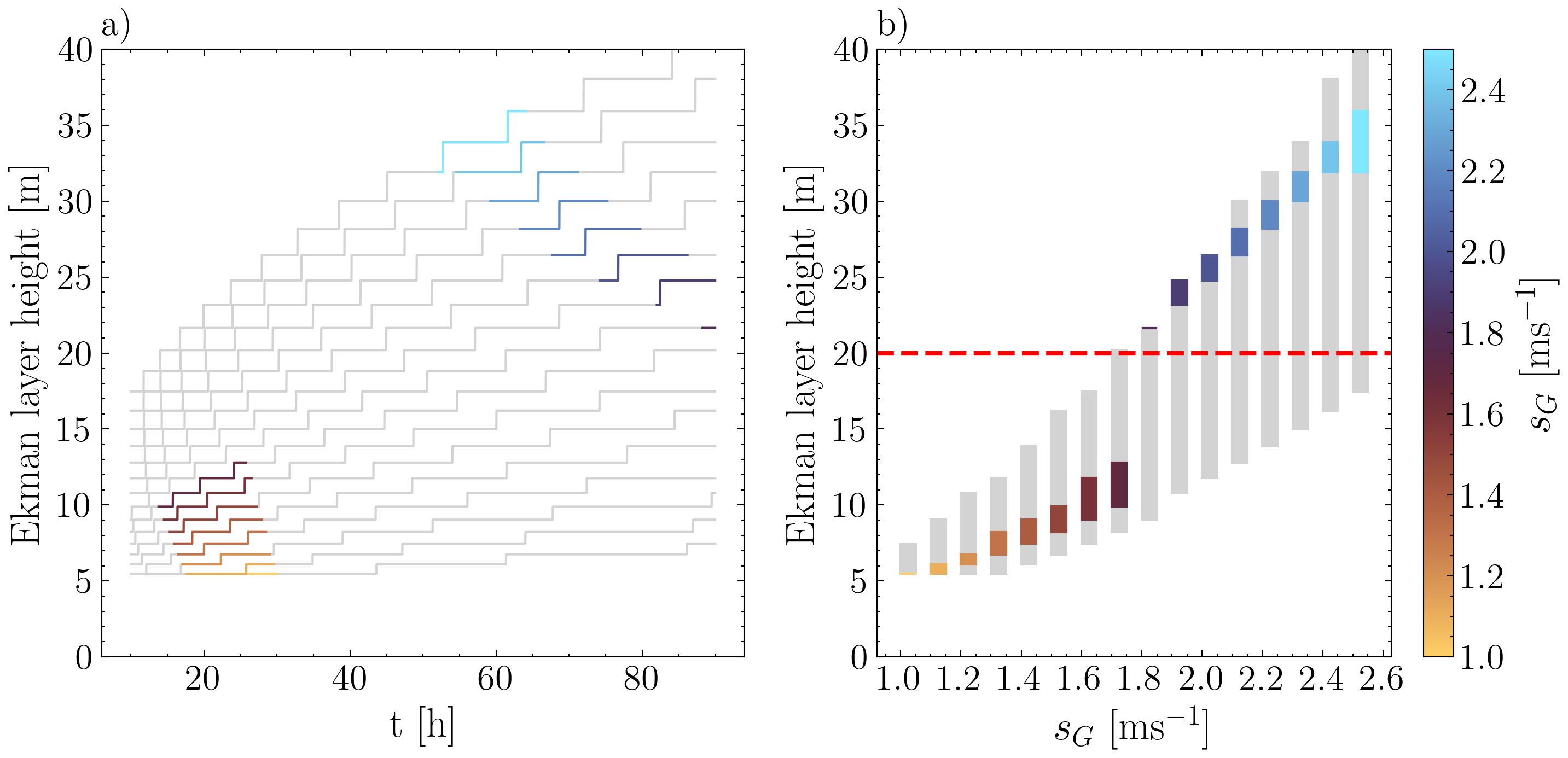}\\
  \caption{a): Ekman layer height plotted over time. Every grey line is a simulation with a different geostrophic wind forcing. The colored sections of the lines mark the location of the quasi-stationary states and the following hour. The colors indicate the value of the geostrophic wind for that particular simulation. b): Plots of the Ekman layer height for all considered $s_G$. The grey boxes show all obtained height values for the whole simulation time. Again, the colored sections are the values of the Ekman layer height at the quasi-stationary state and the following hour, with the colors indicating the $s_G$ value for that particular simulation. The red dashed line in the plot marks $z_{ss}=20$ m. }\label{fig:ekman_layer_height}
\end{figure}

Commonly, a phenomenon where the system undergoes a sudden qualitative change triggered by specific threshold conditions of a relevant parameter, i.e., a bifurcation, is studied with regard to a fixed parameter external to the system, not one of the models internal variables like $u$ that is dependent on height and time. Therefore, in plot \ref{fig:bif_diagram}a $\Delta \theta$ is plotted over $s_G$ instead of $s$ for the same simulation setup as in plot \ref{fig:bif_diagram}b. Again for every simulation, only $s_G$ is changed, and in particular, the initial values are defined as in Section \ref{sec:model}\ref{sec:initial_bound_cond} and not perturbed. Hence, this graph cannot display any backfolding. Indeed, observing backfolding would require at least two distinct simulations, i.e. with different initial conditions, with two different equilibrium states but with the same wind speed. That no backfolding is observed when plotting $\Delta \theta$ over $s_G$ agrees with results from an observational study at Cabauw, Netherlands, by \cite{van_der_linden_local_2017}. They could also not find a critical geostrophic wind speed or a narrow range of geostrophic winds for which sudden or sharp regime transitions were observed. However, they found sharp transitions when turbulence fluxes were related to the instantaneous tower wind speed both observed at a fixed height which is compatible with our results based on a fixed height. To investigate the bistability of the model further, an inherent nonlinearity in the boundary layer dynamics is analyzed by studying the growth of the Ekman layer for increasing wind forcing. The Ekman layer height is estimated as the height where $s$ equals $s_G$. The Ekman layer height for all considered $s_G$ is plotted over time in figure \ref{fig:ekman_layer_height} a). The first 10 hours are considered the initialization period and are therefore not displayed. The step-like structure of the lines can be explained by the non-equidistant grid for the space dimension, $z$. Plot b) shows how the Ekman layer height spread over time evolves with increasing $s_G$. The red horizontal dashed line shows the upper height for $\Delta \theta$, i.e., $z_{ss}=20$~m. The gray boxes represent all $\Delta \theta$ values obtained over the 90 hours of simulation time without the initializing period for every $s_G$ forcing. The quasi-stationary state plus the following 12 hours are colored according to $s_G$ in plots a) and b). Plot b) shows a nearly linear relationship between the Ekman layer height and $s_G$. For $s_G$ less than 1.8 $\mathrm{ms^{-1}}$, the quasi-stationary state is below $z_{ss}$. This correlates with the start of the potential weakly stable regime in the bifurcation diagram (figure \ref{fig:bif_diagram}b). The observed backfolding in figure \ref{fig:bif_diagram}b and the described phenomena in the literature can hence be explained by the fact that the temperature difference is studied for a fixed height, which is below the Ekman layer height for low geostrophic winds and above it for higher wind speeds. If the observation height is very close to the Ekamn layer height, then $\Delta \theta$ is very sensitive to small variations in the Ekman layer height. This can explain the observed backfolding in a wind speed range for which the Ekman layer height is around the measured height. Therefore, this backfolding could be an artifact of the fixed height at which the observations were made and not the result of a sudden rapid shift in the dynamics or an actual sign of bistability.
These results suggest that large-scale forcing, like a change in geostrophic wind, is a main driver for regime transitions in the SBL, i.e., bifurcation-driven transitions, and that the transitions may rather correspond to linear growth of the Ekman layer height and to its height reaching the measurement level. Nonetheless, localized mixing due to small-scale fluctuations and intermittent events can be an additional factor that causes regime transitions, as uncertainty in the parameterization may increase the coupling (shift to weakly stable) or decrease it (shift to very stable). These types of transitions are called noise-induced transitions. As an example, local small-scale wind speed uptake increases shearing, leading to an increase in mixing and resulting in an increase in the thickness of the SBL and further coupling (weakly stable regime) of the boundary layer. Moreover, a burst of turbulence may non-linearly amplify mixing, leading again to further coupling.

\section{Sensitivity analysis}\label{sec:sensitivity_analysis}
To explore the impact of small-scale perturbations on regime transitions and identify the relevant variables' small-scale fluctuations, two approaches are used. First, the wind and temperature tendencies are perturbed to represent localized acceleration/deceleration of the wind or cold/warm air advection. This is achieved by introducing deterministic localized bursts of varying strength and size to the wind and temperature dynamics (see Section \ref{sec:sensitivity_analysis}\ref{sec:perturbed_wind_tempearture}). In part \ref{sec:effect_perturbation_size} of Section \ref{sec:sensitivity_analysis}\ref{sec:perturbed_wind_tempearture}, the influence of the perturbation size is quantified, and in part \ref{sec:effect_perturbation}, four different stability scenarios are analyzed in detail. Opting for a deterministic perturbation facilitates a more straightforward quantification of its effects than a stochastic perturbation. Moreover, computational costs are significantly lower, as a stochastic perturbation would require ensemble runs to quantify its effects. Nevertheless, exploring the effects of stochastically perturbed tendencies is an important avenue for future research. Second, the parameterization of turbulence is perturbed to study the influence of turbulence intermittency. This is realized using a stochastic stability equation that incorporates localized turbulent bursts (see Section \ref{sec:random_stab_func}). 

\subsection{Perturbed model tendencies}\label{sec:perturbed_wind_tempearture}
\subsubsection{Perturbation definition}\label{sec:perturbation_def}
\begin{figure}[t]
  \centering  \noindent\includegraphics[width=0.5\textwidth,angle=0]{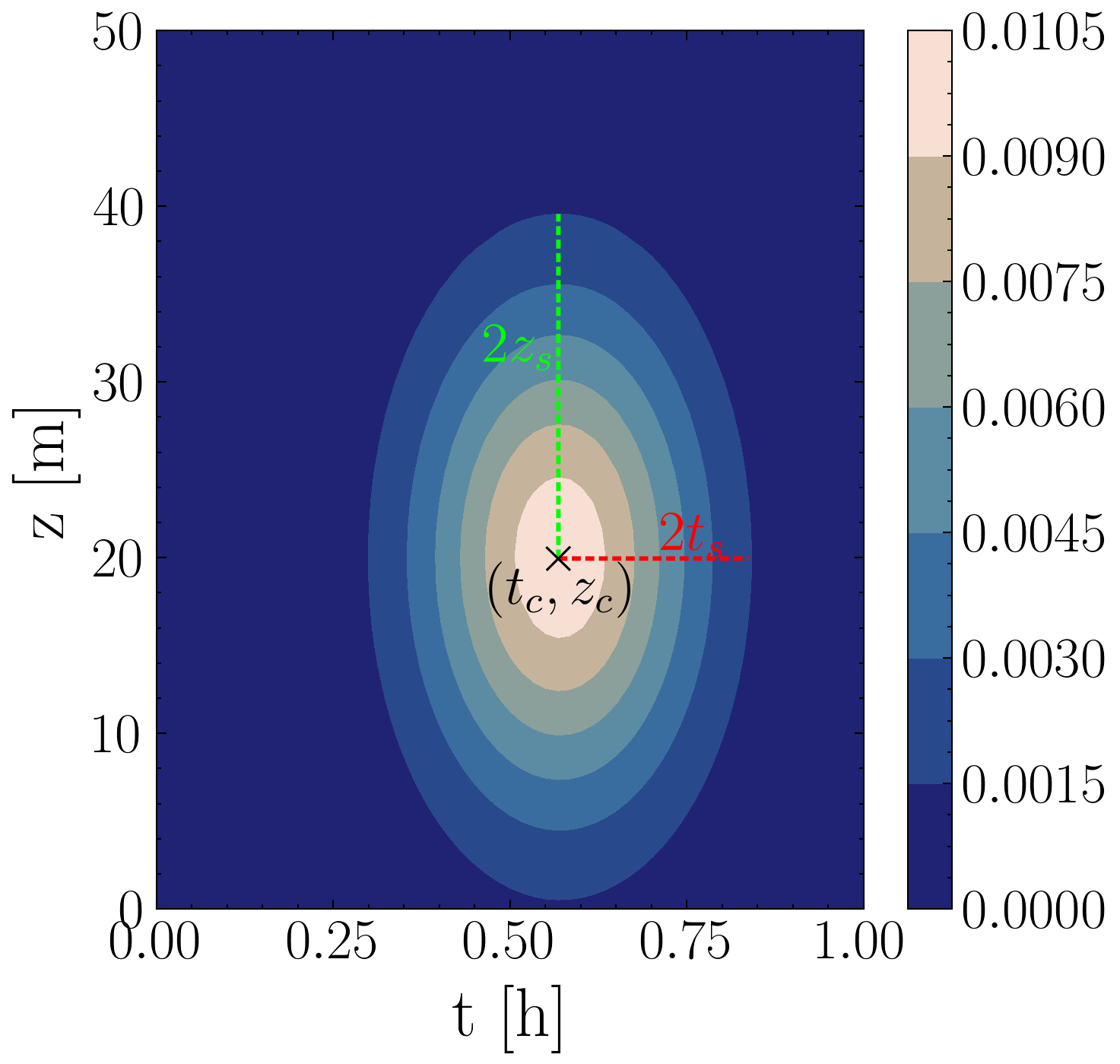}\\
  \caption{Example of a positive perturbation (eq. \eqref{eq:def_perturbation_gaussian}) with maximal perturbation strength, $r$, equal to 0.01, height spread $z_s=10$~m and time spread 
 $t_s=200$ s.}\label{fig:ex_perturbation}
\end{figure}

To perturb the dynamics of the wind and temperature locally in space and time, a 2D Gaussian function is used. This function is defined as 
\begin{equation}
    p(t,z) = r \cdot \exp{\left(-\left[\frac{(t-t_c)^2}{2t_s^2}+\frac{(z-z_c)^2}{2z_s^2}\right]\right)} .\label{eq:def_perturbation_gaussian}
\end{equation}
where $r$ is the amplitude or maximum perturbation strength, $t_c$ and $z_c$ are the centers of the perturbations pulse in time and space, and $t_s$ and $z_s$ control the time and space spread of the perturbation. The actual time and height spread of the perturbation are roughly $4t_s$ and $4z_s$. The unit of $r$ depends on which variable is perturbed. If $u$ is perturbed, the unit of $r$ is $\mathrm{ms^{-2}}$, while for a perturbation of $\theta$, it is $\mathrm{Ks^{-1}}$. The perturbations are localized bursts, limited in height and time and small in magnitude. They fade out over a short time span. Figure \ref{fig:ex_perturbation} shows an example of such a perturbation.

To analyze the influence of small-scale changes in the temperature dynamics and analyze if the localized perturbations can trigger regime transitions, equation \eqref{eq:pde_theta} is perturbed as follows
\begin{align}
    \frac{\partial \tilde{\theta}}{\partial t} &= \frac{\partial}{\partial z}\left(K_h(\phi) \frac{\partial \theta}{\partial z}\right) \pm p(t,z) \nonumber\\
    &= \frac{\partial \theta}{\partial t} \pm p(t,z).\label{eq:pde_theta_perturbed}
\end{align}
The wind speed equation is perturbed similarly to analyze the sensitivity of the model to small-scale changes in the wind dynamics, i.e.,
\begin{align}
    \frac{\partial \tilde{u}}{\partial t} &= f_c \cdot (v - v_G) + \frac{\partial}{\partial z}\left(K_m(\phi) \frac{\partial u}{\partial z}\right) -N_u \pm p(t,z) \nonumber\\
    &= \frac{\partial u}{\partial t} \pm p(t,z).\label{eq:pde_u_perturbed}
\end{align}
If the simulation starts in a very stable regime a negative perturbation is added to $\theta$ and a positive one to $u$ to induce transitions. In contrast, if it starts in the weakly stable regime, a positive perturbation is added to $\theta$ and a negative one to $u$. A negative temperature perturbation locally modifies the temperature gradient, which potentially leads to buoyant forces being locally destabilizing and, as a result, top-down mixing. If the mixing becomes sufficiently strong it results in a coupling of the layers. An increase in mixing is also expected for a positive wind perturbation. On the other hand, a positive temperature perturbation is expected to increase the temperature gradient, which stabilizes the buoyant forces and thereby inhibits mixing. Moreover, a negative wind perturbation is expected to reduce the wind speed and hence shearing, which results in a reduction of mixing.”

\subsubsection{Effect of the perturbation size}\label{sec:effect_perturbation_size}
\begin{figure}[t]
  \noindent\includegraphics[width=\textwidth,angle=0]{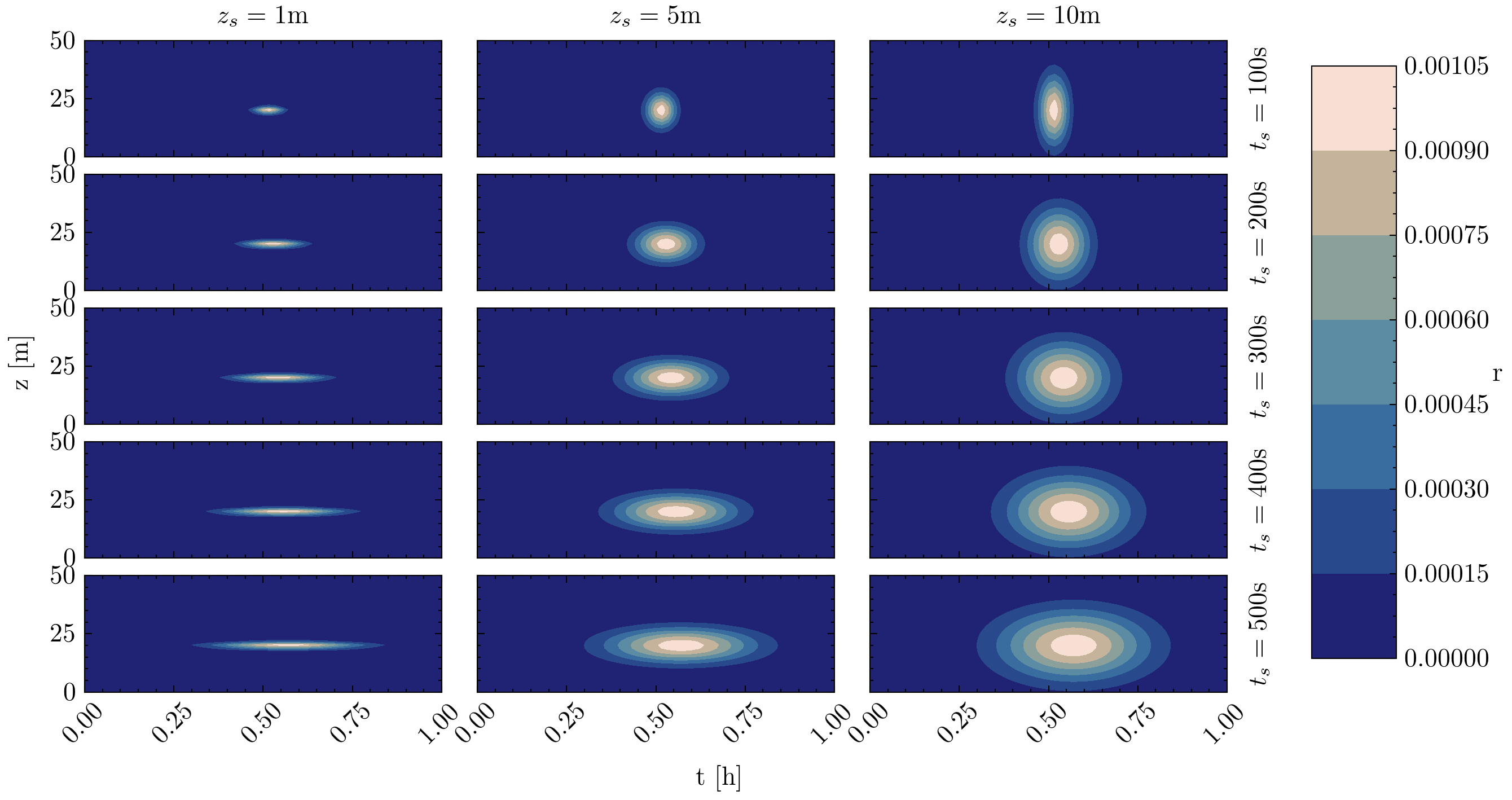}\\
  \caption{This figure gives an overview of all considered perturbations. In each column, the height spread, $z_s$, is constant, while in every row, the time spread, $t_s$, is constant. That means from left to right, the spread in height increases, and from top to bottom, the time spread increases. The strength of the perturbation is indicated by color. The maximal perturbation strength is equal to 0.001 for all plotted perturbations.}\label{fig:perturbation_size_comp}
\end{figure}

\begin{figure}[t]
  \noindent\includegraphics[width=\textwidth,angle=0]{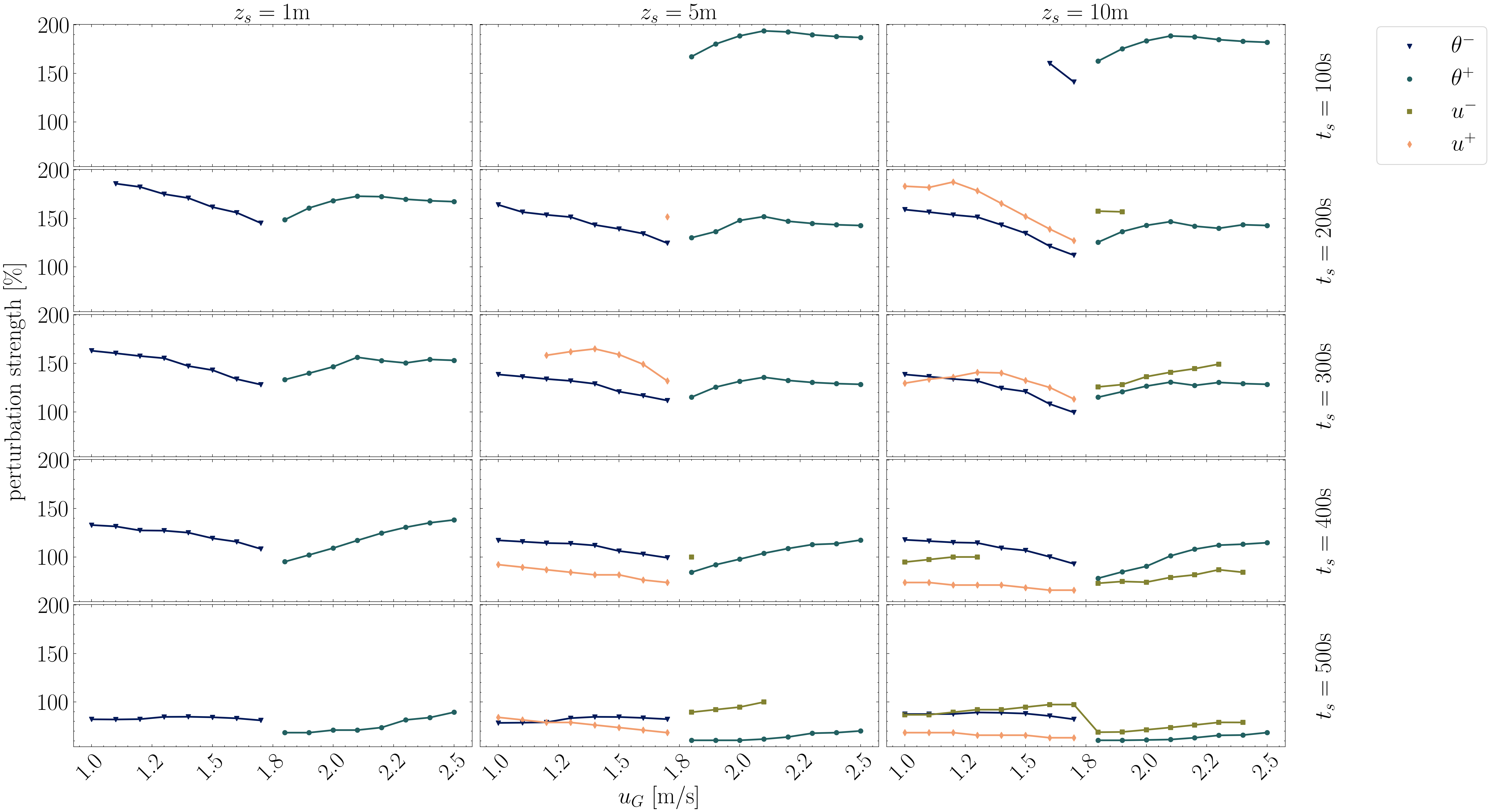}\\
  \caption{This figure gives an overview of the results of the sensitivity analysis for all considered perturbations (see figure \ref{fig:perturbation_size_comp}). In each column, the height spread, $z_s$, is constant, while in every row, the time spread, $t_s$, is constant. That means from left to right, the spread in height increases, and from top to bottom, the time spread increases. The plots show the minimal perturbation strength for which that simulation includes a transition. The perturbation strength is here given in relation to the standard deviation of the simulation's first 15 minutes, which are not perturbed.}\label{fig:sensitivity_analysis_u_v_theta}
\end{figure}

The perturbation is characterized by four parameters that control its location and spread in space and time. Different values for the time and space spread are investigated to study the effect of the perturbation size. The center of the perturbation is fixed at 20~m to allow easy comparison. This height is chosen as the quasi-stationary state was determined at this height. The time center of the perturbation is between 30 and 35~min for all simulations. By choosing this time window for the perturbation center, a range of time spreads can be tested without the perturbation extending beyond the initial condition. The maximal time duration of the perturbation which is considered is 33~min ($t_s=500$~s), and the maximal space extension is 40 m ($z_s=10$~m), which is in the realm of submesomotions \citep{mahrt_stably_2014, vercauteren_investigation_2016, boyko_multiscale_2021}. The smallest extension considered is 7~min ($t_s=100$~s) in time and 4~m ($z_s=1$~m) in space. Figure \ref{fig:perturbation_size_comp} shows all considered perturbations, and figure \ref{fig:sensitivity_analysis_u_v_theta} shows the results of the corresponding sensitivity analysis. The sensitivity analysis determines the minimal perturbation strength for every combination of geostrophic wind, perturbation type, and size for which at least one transition takes place in the 12~h simulation time. Here, perturbation strength refers to the relation between $r$, i.e. the maximal perturbation strength, and the standard deviation of the corresponding variable ($u$ or $\theta$) for the first 15 minutes of the simulation. This period is chosen as the simulations are not perturbed during this period for all perturbation sizes. As the tested $r$ values and the standard deviations have different scales, both are normalized. A min-max normalization is applied to transform the values such that the minimum value becomes 0, the maximum value 1, and all other values are linearly scaled to a 0 to 1 range. All normalized values are denoted with a bar in the description of the following procedure to determine the perturbation strength:
\begin{enumerate}
    \item All tested $r$ values are normalized, i.e. $\overline{R} = \left[\overline{r_1},\dotsc,\overline{r_k}\right]$. 
    \item The normalized $r$ value for the mth simulation is chosen, i.e. $\overline{r_m}$.
    \item The time standard deviation of $X_m$, where $X_m$ is either $u$ or $\theta$ from the mth simulation, is calculated for all heights, i.e. $\sigma_m(z) = \left[\sigma(X_m(z_0,t)),\dotsc,\sigma(X_m (z_H,t))\right]$.
    \item The standard deviation vector is normalized, i.e. $\overline{\sigma_m(z)}$.
    \item The normalized standard deviation at the height of 20~m is chosen, i.e. $\overline{\sigma_m(20~m)}$.
    \item The perturbation strength is then calculated as $\overline{r_m}/\overline{\sigma_m(20~m)}\cdot 100$.
\end{enumerate}
This is procedure is repeated for all simulations.

For this study, $u_G$ values between 1.0 and 2.5 $\mathrm{ms^{-1}}$ and $r$ values between 0 and 0.019 were tested. For the smallest perturbation, no simulation included transitions. Therefore, no smaller perturbations were considered for this study. 

As expected, negative perturbations added to the temperature dynamics cause regime transitions if the simulations are started in the very stable regime ($u_G\leq 1.7\;\mathrm{ms^{-1}}$) while a positive perturbation causes transitions for simulation started in the weakly stable regime ($u_G\geq 1.8 \;\mathrm{ms^{-1}}$) for nearly all perturbation sizes. This agrees with physical intuition as the inversion strength increases if warm air is advected over a colder surface, while cold air advection decreases the inversion strength if the advection is close to the surface. A higher perturbation strength would most likely yield the expected results in the few cases where this does not seem to hold. However, for the parameters used in this study, a perturbation height spread of 4~m, i.e. $z_s=1$~m, is insufficient to induce transitions (see top left panel of figure \ref{fig:sensitivity_analysis_u_v_theta}). The results of the perturbed temperature dynamics show a noticeable difference when $t_s$ is changed, whereas a change in $z_s$ does not have as pronounced an effect. Unsurprisingly, increasing $t_s$ leads to a decrease in the minimum perturbation strength required to cause transitions.

Analogously, positive perturbations added to the wind dynamics cause regime transitions if the simulations were started in the very stable regime ($u_G\leq 1.7\; \mathrm{ms^{-1}}$) while a negative perturbation causes transitions for simulation started in the weakly stable regime ($u_G\geq 1.8 \;\mathrm{ms^{-1}}$) if the perturbation was large enough. The decrease in stability after an increase in wind speed can be explained by the fact that an increase in wind speed results in more shearing and, thereby, more mixing, ultimately leading to further coupling. For the here used parameters, a total perturbation time spread of 400~s, i.e., $\approx$7~min, or a total height spread of 4~m, is insufficient to get transitions related to perturbed wind dynamics. Contrary to the temperature dynamics for the perturbed wind dynamics, there is a noticeable difference between the results if $z_s$ or $t_s$ is changed. Unsurprisingly, if $t_s$ increases, the minimal perturbation strength to cause transitions decreases. The same holds for an increase in $z_s$, just less drastically. Surprisingly, there are transitions from very to weakly stable for a negative wind perturbation. This can be explained by a wind direction change and is discussed in detail in the following section.

\subsubsection{Effect of perturbation in detail}\label{sec:effect_perturbation}
In this section, the focus is on four different stability scenarios:
\begin{enumerate}
    \myitem{S1.}\label{item:stability_scenario_1} $u_G = 1.0\ \mathrm{ms^{-1}}$: very stable regime, 
    \myitem{S2.}\label{item:stability_scenario_2} $u_G = 1.7\ \mathrm{ms^{-1}}$: transition wind speed of very stable regime,
    \myitem{S3.}\label{item:stability_scenario_3} $u_G = 1.8\ \mathrm{ms^{-1}}$: transition wind speed of weakly stable regime, and
    \myitem{S4.}\label{item:stability_scenario_4} $u_G = 2.5\ \mathrm{ms^{-1}}$: weakly stable regime.
\end{enumerate}

\begin{figure}[t]
  \noindent\includegraphics[width=\textwidth,angle=0]{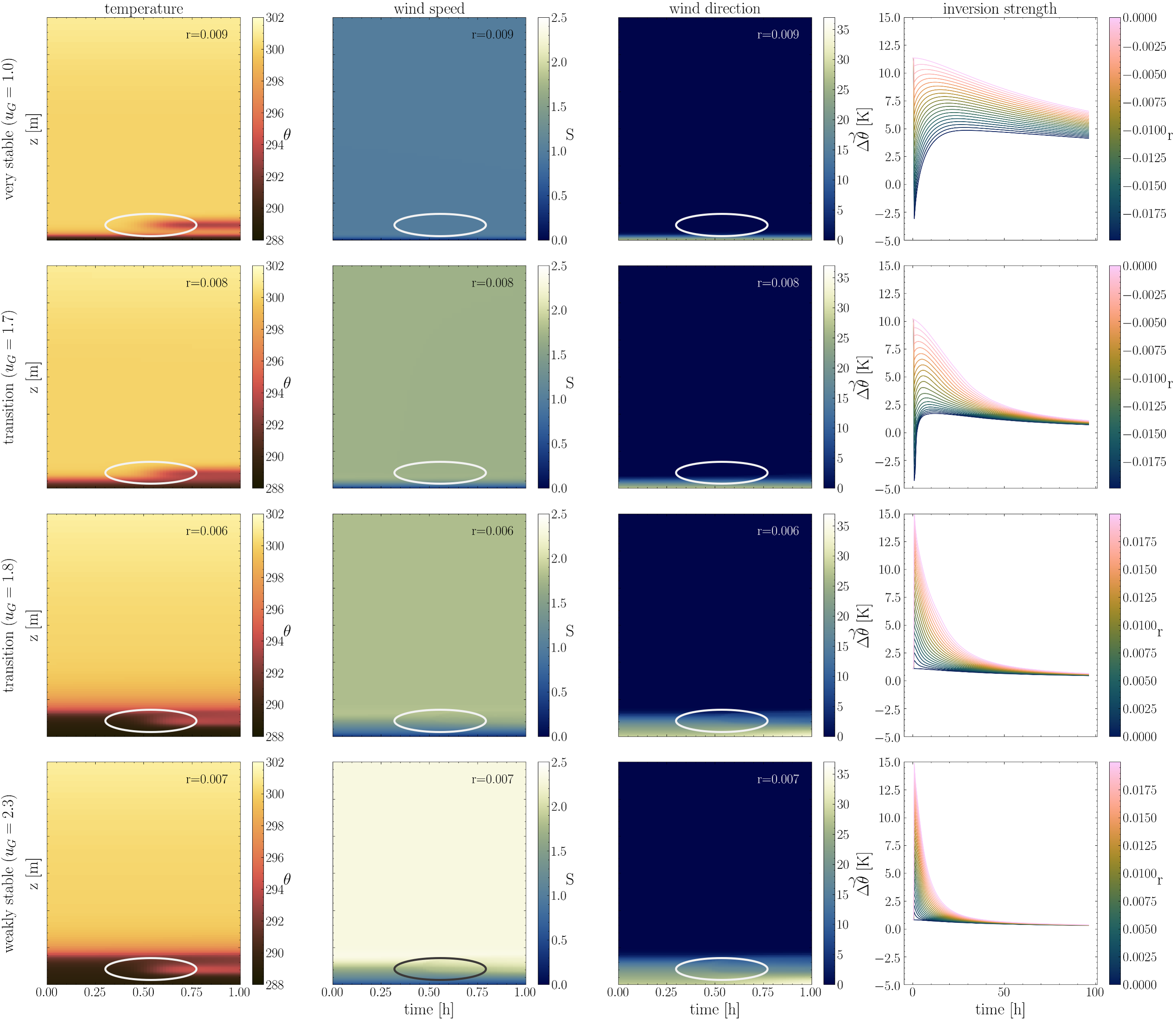}\\
  \caption{Examples of simulations with perturbation added to the temperature dynamics. The plots show the simulation output of the temperature (first column), wind speed (second column), and wind direction (third column) over space and time. The ovals indicate the location of the perturbation. The right column shows the temperature inversion strength plots for a range of perturbation strengths, $r$, indicated by color. The parameters for the perturbation size are $z_s=5$ m and $t_s=300$ s. The geostrophic wind and the perturbation strength are the same in every row. They are given in the title of each plot. Every row corresponds to one of the four stability scenarios (\ref{item:stability_scenario_1} - \ref{item:stability_scenario_4}). In the first/last row, the simulation is initialized in the very/weakly stable regime outside the bifurcation region. In the second/third, the simulation starts in the very/weakly stable regime but at the transition wind speed (see figure \ref{fig:bif_diagram}b). }\label{fig:perturbation_impact_theta}
\end{figure}
\clearpage 

Figure \ref{fig:perturbation_impact_theta} shows the impact of the temperature perturbation on the potential temperature, $\theta$, the wind speed $S$, the wind direction $\gamma$ and the inversion strength $\Delta \theta$. Each row corresponds to one of the four scenarios. The perturbation strength is chosen according to the sensitivity analysis performed in the previous section. The negative and positive perturbations are visible in the plots of $\theta$, while the wind speed and direction do not change significantly. Therefore, the temperature perturbation impacted mainly the temperature and effectively represents the advection of cold or warm air. The plots for the temperature inversion strength (right column in figure \ref{fig:perturbation_impact_theta}) show that for all scenarios, a transition from one regime to another occurs if the perturbation strength is high enough. To reiterate, a transition occurs when the temperature difference crosses the 5 K threshold. Unsurprisingly, the stronger the perturbation is, the stronger the resulting temperature difference is directly after the perturbation. In fact, for all simulations, a change of roughly 13 K of the temperature difference in less than one hour can be observed for the highest perturbation strength. This holds for simulations where $u_G$ is smaller and equal to the transition wind speed for both the weakly and the very stable regime. For the first scenario, for the strongest perturbations, the system quickly transitions but then returns to a state between the very and weakly stable one, i.e., $\Delta \theta$ equals roughly 5~K. However, for the second scenario, the system equilibrates in the weakly stable regime after the transition from very stable to weakly stable conditions, even for a simulation time of 96~h. It shall be noted that the system transitions in this scenario without a perturbation. This unperturbed transition occurs in the span of roughly 40 hours, while in the simulations with perturbations it occurs in less than one hour. For all other scenarios, while the transition occurs rapidly after the perturbation, the system shows a strong tendency to return to its original state for all tested perturbation strengths. This holds especially for scenario 4, where the return to the weakly stable state occurs even for the strongest perturbation after less than 7 hours.

\begin{figure}[t]
  \noindent\includegraphics[width=\textwidth,angle=0]{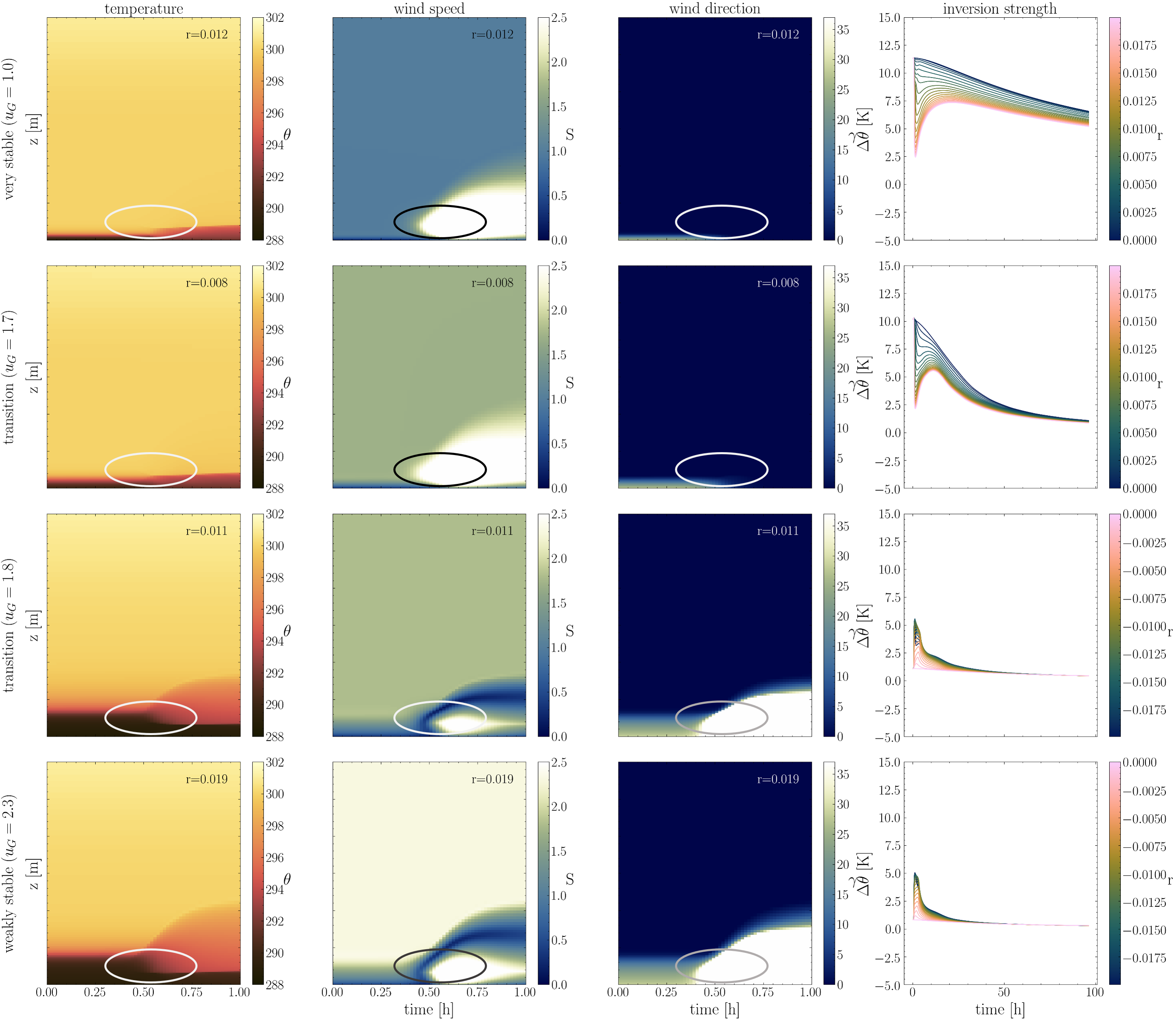}\\
  \caption{Examples of simulations with perturbation added to the wind dynamics. The plots show the simulation output of the temperature (first column), wind speed (second column), and wind direction (third column) over space and time. The ovals indicate the location of the perturbation. The right column shows the temperature inversion strength plots for a range of perturbation strengths, $r$, indicated by color. The parameters for the perturbation size are $z_s=10$ m and $t_s=300$ s. The geostrophic wind and the perturbation strength are the same in every row. They are given in the title of each plot. Every row corresponds to one of the four stability scenarios (\ref{item:stability_scenario_1} -- \ref{item:stability_scenario_4}). In the first/last row, the simulation is initialized in the very/weakly stable regime outside the bifurcation region. In the second/third, the simulation starts in the very/weakly stable regime but at the transition wind speed (see figure \ref{fig:bif_diagram}).}\label{fig:perturbation_impact_u}
\end{figure}
\clearpage

Figure \ref{fig:perturbation_impact_u} is the same figure as figure \ref{fig:perturbation_impact_theta} but with wind velocity perturbation instead of temperature perturbation. For the positive perturbation, i.e., scenarios 1 and 2, the wind speed plots look as expected. At the center of the perturbation, the wind speed is the highest and then decreases in the space and time dimension. The wind direction changes slightly but not drastically. Therefore, in these two cases, adding a perturbation to $u$ causes mainly a perturbation in the wind speed. Similar to the perturbed temperature dynamics, if the simulations are started in the very stable regime, transitions are observed if the perturbation strength is high enough for both stability scenarios. However, the maximal temperature difference is smaller for perturbed wind than the perturbed temperature. The maximal decrease is roughly 8K instead of 13K.

Contrarily to the positive perturbations, the negative perturbations significantly impact the wind direction. In fact, at the center of the perturbation, the wind turns with nearly 180$^{\circ}$. This also explains why the wind speed increases at the center of the perturbation, even though a negative perturbation was added. The wind velocity component $u$ decreases so much that it gets negative, which is equivalent to a wind direction change of 180 $^{\circ}$. The reversal of the wind direction explains the phenomena discovered in the previous section: a negative wind perturbation causes very to weakly stable transitions for some perturbations. This is not the intended behavior, and several approaches to prevent this have been tried. But all approaches led to strong numerical instabilities. Hence, the results obtained with a negative wind perturbation must be treated with caution. Nonetheless, the temperature inversion strength plots show increased stability after the negative perturbation for all tested perturbation strengths. In contrast to the perturbed temperature dynamics, the regime threshold is barely crossed for the highest perturbation strengths. Moreover, there are no persistent transitions into the very stable regime ($\Delta \theta >5$~K). 

In conclusion, the combined results for both perturbation types indicate that small-scale phenomena can drive persistent regime transitions from very to weakly stable regimes. In contrast, the phenomena tested here are insufficient to cause persistent weakly to very stable regime transitions. Therefore, it is especially relevant to account for small-scale fluctuations in the wind and temperature dynamics in weather and climate models to achieve accurate regime transition statistics for very to weakly stable transitions.

\subsection{Changes in the system's stability}\label{sec:ekman_layer_perturbed}
\begin{figure}[t]
  \noindent\includegraphics[width=\textwidth,angle=0]{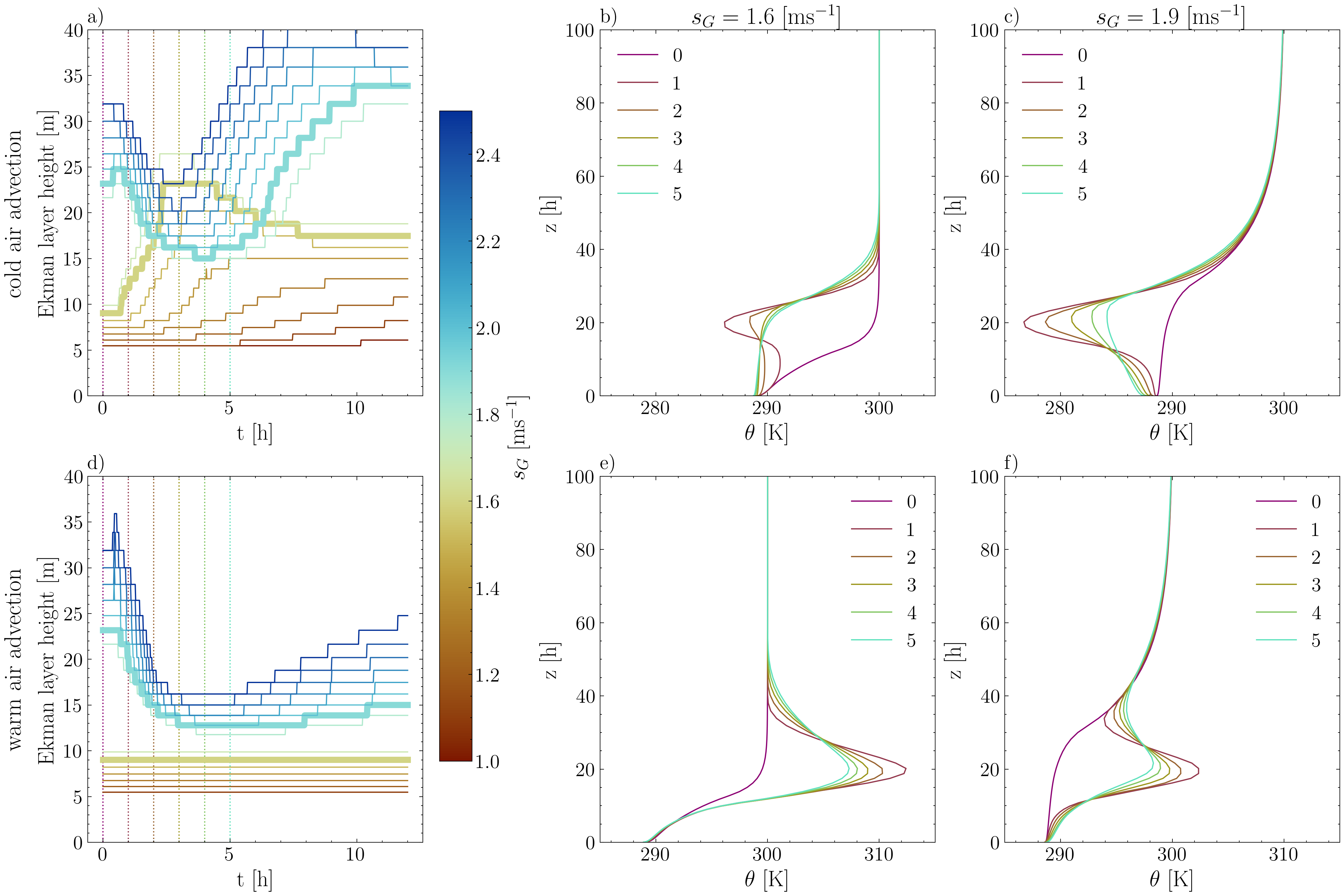}\\
  \caption{First row: Results with negatively perturbed temperature dynamics (cold air advection). Second row: positively perturbed temperature dynamics (warm air advection). First column: Ekman layer height plotted over time. Every line is a simulation with a different geostrophic wind forcing, indicated by the color. The two thicker lines are for $s_G=1.6$ and $1.9\ \mathrm{ms^{-1}}$. Second column: Temperature profiles for $u_G=1.6\ \mathrm{ms^{-1}}$ and six different simulation time points (marked as vertical dotted lines in column one). Third column: Same as second column, just for $u_G=1.9\ \mathrm{ms^{-1}}$.}\label{fig:ekman_layer_height_rand}
\end{figure}

After studying the Ekman layer height in Section \ref{sec:model}\ref{sec:bifurcation_analysis} for the purely deterministic model, we now study the Ekman layer height of a perturbed version of the model to analyze if there is an inherent nonlinearity in the boundary layer dynamics in the sense that the growth of the Ekman layer becomes nonlinear for a certain range of wind speed. For this, the model with a temporally and spatially localised perturbation added to the temperature dynamics (eq.~\eqref{eq:pde_theta_perturbed}) is initialized at the quasi-stationary state of the deterministic model and run for 12 hours. The same perturbation as in Section \ref{sec:sensitivity_analysis}\ref{sec:perturbed_wind_tempearture} is used. The perturbation is a small, localized burst of small magnitude, which is added in the first hour of the simulation and then fades out in that hour. The center of the perturbation is at 20~m. The results are shown in figure \ref{fig:ekman_layer_height_rand}. For plot a) to c), a negative temperature perturbation is added, equating to cold air's advection. As the perturbation is added at a fixed height for geostrophic winds less than $1.8~\mathrm{ms^{-1}}$, the center of the perturbation is above the Ekman layer height, while for higher wind speeds, it is below. 

Figure \ref{fig:ekman_layer_height_rand} c) shows a plot of the temperature profiles for the perturbed solution for $u_G = 1.6~\mathrm{ms^{-1}}$. The different lines correspond to different simulation time points. As $u_G<1.8~\mathrm{ms^{-1}}$, the center of the perturbation is above the height of the Ekman layer. As a result of the perturbation, the temperature decreases at roughly 20~m, which is the center of the perturbation and higher than the Ekman layer at the simulation's start. Hence, the perturbation causes cooling above the Ekman layer, creating a strong negative vertical temperature gradient between the inversion layer and the cooled layer above. This causes a buoyant instability and thus an increase in mixing and, thereby, an increase in the Ekman layer height until a new equilibrium state is reached. In plot c), the temperature profiles for $u_G = 1.9~\mathrm{ms^{-1}}$ are shown. Here, the center of the perturbation is below the height of the Ekman layer. The negative perturbation causes a capping inversion to form between 20~m and 40~m, partially below the Ekman layer height, effectively reducing the height of the mixed layer and thus reducing the Ekman layer height. But as a counter-movement, a convective state emerges in the first 20~m, which allows the mixing to take up more space again and thereby increases the Ekman layer height again. As the model nonlinearly amplifies this process, the Ekman layer height rises above the original height from the start of the simulation and equilibrates in this state.

For plots d) to f), a positive temperature perturbation is added, equating to warm air's advection. Again, for geostrophic winds less than $1.8~\mathrm{ms^{-1}}$, the perturbation is added above the Ekman layer height, while for higher wind speeds, the perturbation is added below. Plot e), $u_G = 1.6~\mathrm{ms^{-1}}$, shows when the perturbation is added above the Ekman layer and for low geostrophic winds, it does not affect the Ekman layer height. While the perturbation causes the temperature to increase at roughly 20~m, this is above the Ekman layer height, and hence, the inversion layer is only strengthened, and no change in mixing occurs, which would result in a change in the Ekman layer. Above the Ekman layer height, i.e., between 20 and 40~m, a detached layer forms, which is convectively unstable but does not induce top-down mixing but rather propagates mixing into higher layers.

For $u_G = 1.9~\mathrm{ms^{-1}}$, the perturbation center is below the Ekman layer height as seen in plot f). Hence, even though the perturbation has the same effect as for $u_G = 1.6~\mathrm{ms^{-1}}$, namely the increase in temperature, the strengthening of the inversion further reduces mixing, resulting in a reduction of the Ekman layer height until it starts to return to its original state.

As a low Ekman layer correlates with a strong temperature inversion, i.e., the very stable regime, and a higher one with a weak temperature inversion, i.e., the weakly stable regime, the results above support the observation from the previous section that small-scale perturbations have a long-lasting impact for very to weakly stable transitions. This results from the top-down mixing event recoupling the previously decoupled low SBL to get a much deeper SBL. For weakly to very stable transitions, small-scale perturbations were added at a height relatively close to the surface as commonly observed in field studies \citep{sun_turbulence_2012, lan_turbulence_2022}, but they only have a transient impact.

\subsection{Perturbed parameterization}\label{sec:random_stab_func}
As the effect of perturbed tendencies was studied in the previous section (Sect. \ref{sec:sensitivity_analysis}) in this section, the focus lies on the effect of a perturbed parameterization. This is achieved by using a randomized stability function.
The set of stability functions proposed in the literature is large. While developed with the goal to be universal, many slightly different functions have been derived based on field studies (examples are given in \cite{rodrigo_investigation_2013}). They introduce different types of model errors or uncertainty. This section uses a random stability equation, which was initially developed in \cite{boyko_stochastic_2023} to represent model uncertainty in the turbulence closure. The authors sought to capture unsteady mixing, potentially linked to turbulence intermittency and unresolved fluid motions. Their strategy involves incorporating stochastic parameterization by introducing randomness into the stability correction utilized in the traditional Monin–Obhukov similarity theory. They developed a data-driven stability correction equation using a statistical model-based clustering technique \citep{boyko_statistical_2022}. The resulting stochastic stability equation incorporates random perturbations to address localized intermittent turbulence bursts. This approach considers stochastic mixing effects from unresolved processes.

The stochastic stability equation is briefly introduced here, but more details are given in \cite{boyko_stochastic_2023}. The equation is defined as
\begin{equation}
    \rho(t, Ri) = \phi(Ri) \text{sig}(z)+\phi_s(t,Ri)(1-\text{sig}(z))\label{eq:stoch_stab_func}
\end{equation}
with $\phi(Ri)=1+12Ri$ being the deterministic short-tail stability function, which is defined in section \ref{sec:model}\ref{sec:model_description} (eq. \ref{eq:short_tail_stab_func}). The function $\text{sig}(z)=\frac{1}{1+\exp(-k_c(z-z_m))}$ is the logistic sigmoid function where $z_m$ is the sigmoid’s midpoint, and $k_c = 0.1$ controls the sharpness of the transition from $\phi$ to $\phi_s$ at height $z_m$. The stochastic part of equation \eqref{eq:stoch_stab_func}, $\phi_s(t,Ri)$, is given as the solution to the stochastic differential equation
\begin{equation}
    d\phi_s(t,Ri)=[1-\Lambda(Ri)\phi_s-\Upsilon(Ri)\phi_s^2]dt+\Sigma(Ri)\phi_sdW(t), \quad \phi_s(0)=\phi_{s0}\;.
\end{equation}
where $W(t)$ is a Wiener process. Using a model-based clustering method, \cite{boyko_stochastic_2023} obtain the following data-driven scaling functions
\begin{align}
    \Lambda(Ri) &= 9.3212\tanh[0.9088\log_{10}(Ri)- 0.0738]+ 8.3220\\
    \Upsilon(Ri) &= 10^{0.4294\log_{10}(Ri)+ 0.1749}\\
    \Sigma(Ri) &= 10^{0.8069\tanh[0.6044\log_{10}(Ri)-0.8368]+\sigma_s}.
\end{align} 
The parameter values of these functions were estimated with a regression analysis of the FLOSS II dataset \citep{boyko_stochastic_2023}.

The stochastic stability equation adjusts the turbulent mixing length based on the dimensionless buoyant-to-shear forces ratio. The previous section studied the buoyancy perturbations due to temperature perturbations and shear perturbations stemming from wind perturbations separately. The stochastic stability equation is a pragmatic way to examine and model the combined effect of buoyancy and shear perturbations.

In figure \ref{fig:stab_funcs}, the stochastic stability equation with a high noise value, i.e., $\sigma_s=1$, is compared to the deterministic short- and long-tail stability functions. The plot shows that the deterministic stability functions go to infinity for higher Richardson numbers, resulting in the mixing length going to zero. In contrast, the stochastic stability equation takes values that decrease again at a Richardson number of roughly 1, thereby allowing turbulence to exist even for higher Richardson numbers.

To study the impact of using the randomized stability equation, the same model as in section \ref{sec:sensitivity_analysis}\ref{sec:perturbed_wind_tempearture} (i.e., equations \eqref{eq:pde_u} to \eqref{eq:pde_theta_g}) is used. The noise intensity is controlled by $\sigma_s$, which is independent of the stratification. Hence, this parameter is chosen for the sensitivity study. 
\begin{figure}[t]
  \noindent\includegraphics[width=\textwidth,angle=0]{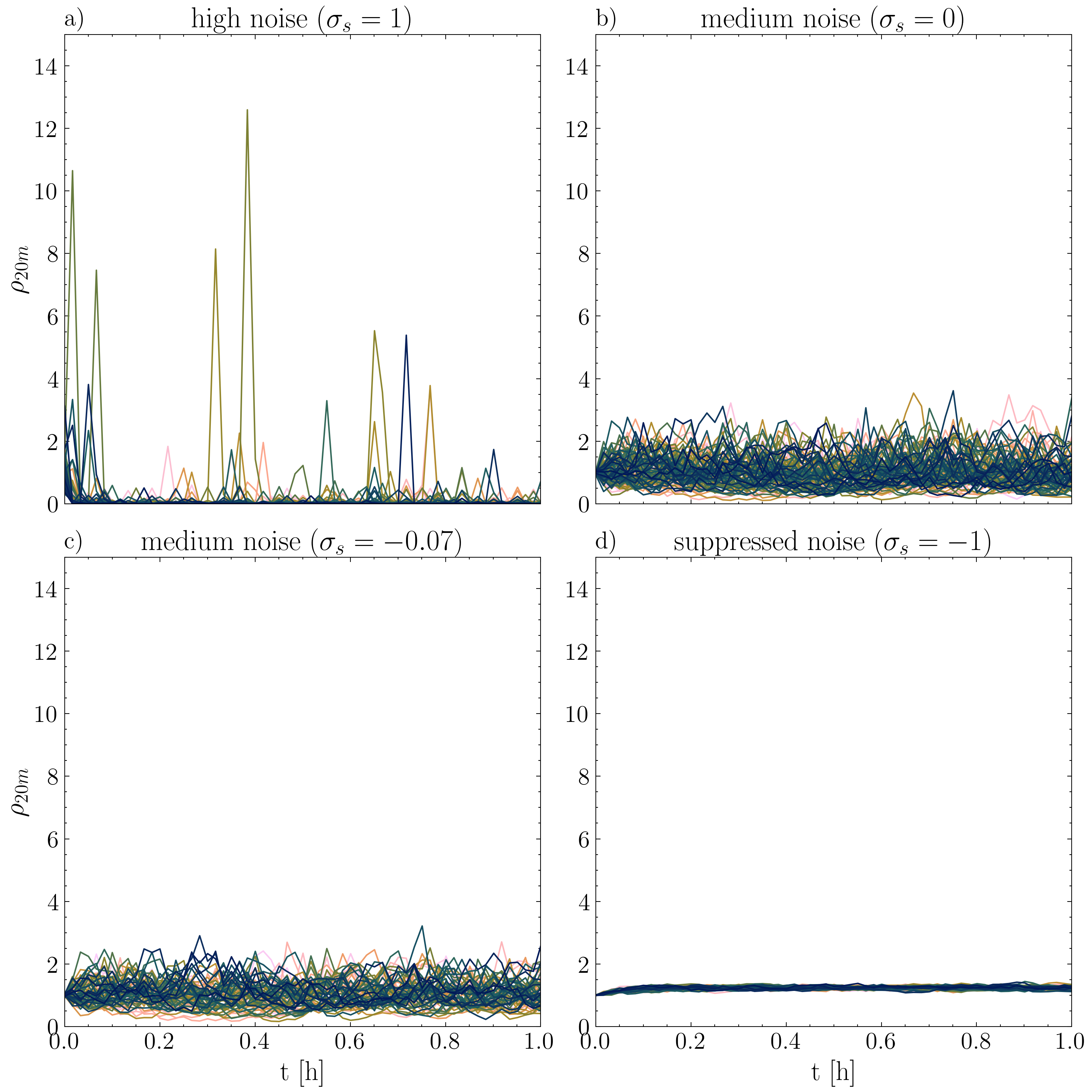}\\
  \caption{Examples of the stochastic stability equation (eq. \ref{eq:stoch_stab_func}) at 20~m with varying noise strengths: a) $\sigma_s=1$, b) $\sigma_s=0$, c) $\sigma_s=-0.07$ and d) $\sigma_s=-1$. Each plot shows 100 examples of $\rho$. The geostrophic wind is $1.0~\mathrm{ms^{-1}}$ for all of the simulations.}\label{fig:stab_func_examples}
\end{figure}
\clearpage
Figure \ref{fig:stab_func_examples} shows examples of the stochastic stability equation for four different $\sigma_s$. Due to the stochasticity and temporal memory in the definition of $\rho$, the stability correction creates temporally stronger mixing events (see, e.g., plot a)). 
\begin{figure}[t]
  \noindent\includegraphics[width=\textwidth,angle=0]{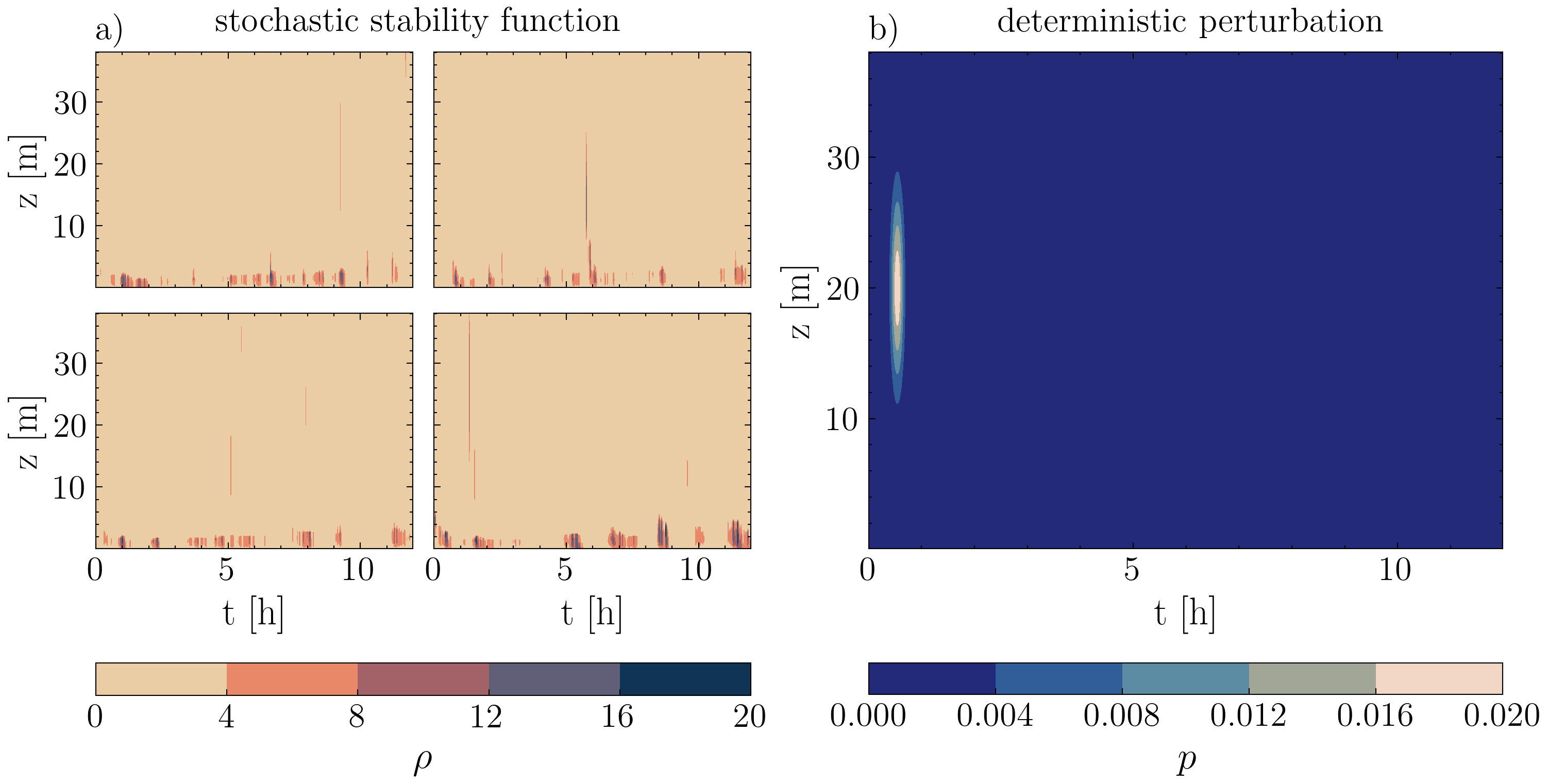}\\
  \caption{Comparison of a) the stochastic stability equation (eq. \ref{eq:stoch_stab_func}) and b) the deterministic perturbation (eq. \ref{eq:def_perturbation_gaussian}). Figure a) shows four examples of the stochastic stability equation with a noise intensity of $\sigma_s=1$. The values are generated with $s_G=1 \ \mathrm{ms^{-1}}$.}\label{fig:stoch_stab_vs_det_perturbation}
\end{figure}
However, the plots show that a decrease in $\sigma_s$ results in a decrease in the variability of $\rho$ until for $\sigma_s=-1$ the noise is nearly fully suppressed. The mixing events occur multiple times during the simulation period and are mostly close to the surface, as shown in figure \ref{fig:stoch_stab_vs_det_perturbation} a). This differs from the perturbation used in section \ref{sec:sensitivity_analysis}, where the perturbation was a singular pulse with a center at 20~m and a height spread of multiple meters (see figure \ref{fig:stoch_stab_vs_det_perturbation} b)). Hence, the stochastic stability equation can be seen as a perturbation strategy with increased complexity compared to the previous purely deterministic perturbation.

The study by \cite{boyko_stochastic_2023} analyzes results with three different values for $\sigma_s$. Here, the same values are used for the sensitivity study, and a high-noise scenario is added. The following four noise scenarios are studied:
\begin{enumerate}
    \myitem{N1.}\label{item:noise_scenario_1} $\sigma_s=1$: high noise level,
    \myitem{N2.}\label{item:noise_scenario_2} $\sigma_s=0$: medium noise level, estimated based on FLOSS II dataset \citep{boyko_stochastic_2023},
    \myitem{N3.}\label{item:noise_scenario_3} $\sigma_s=-0.07$: medium noise level, adjusted noise level for FLOSS II dataset \citep{boyko_stochastic_2023} and
    \myitem{N4.}\label{item:noise_scenario_4} $\sigma_s=-1$: suppressed noise level.
\end{enumerate}
In addition, the same stability scenarios as in section \ref{sec:sensitivity_analysis}\ref{sec:perturbed_wind_tempearture} are used:
\begin{enumerate}
    \myitem{S1.}\label{item:stability_scenario_sf1} $u_G = 1.0\ \mathrm{ms^{-1}}$: very stable regime,
    \myitem{S2.}\label{item:stability_scenario_sf2} $u_G = 1.7\ \mathrm{ms^{-1}}$: transition wind speed of very stable regime,
    \myitem{S3.}\label{item:stability_scenario_sf3} $u_G = 1.8\ \mathrm{ms^{-1}}$: transition wind speed of weakly stable regime, and
    \myitem{S4.}\label{item:stability_scenario_sf4} $u_G = 2.5\ \mathrm{ms^{-1}}$: weakly stable regime.
\end{enumerate}
\begin{figure}[t]
  \centering\noindent\includegraphics[width=0.9\textwidth,angle=0]{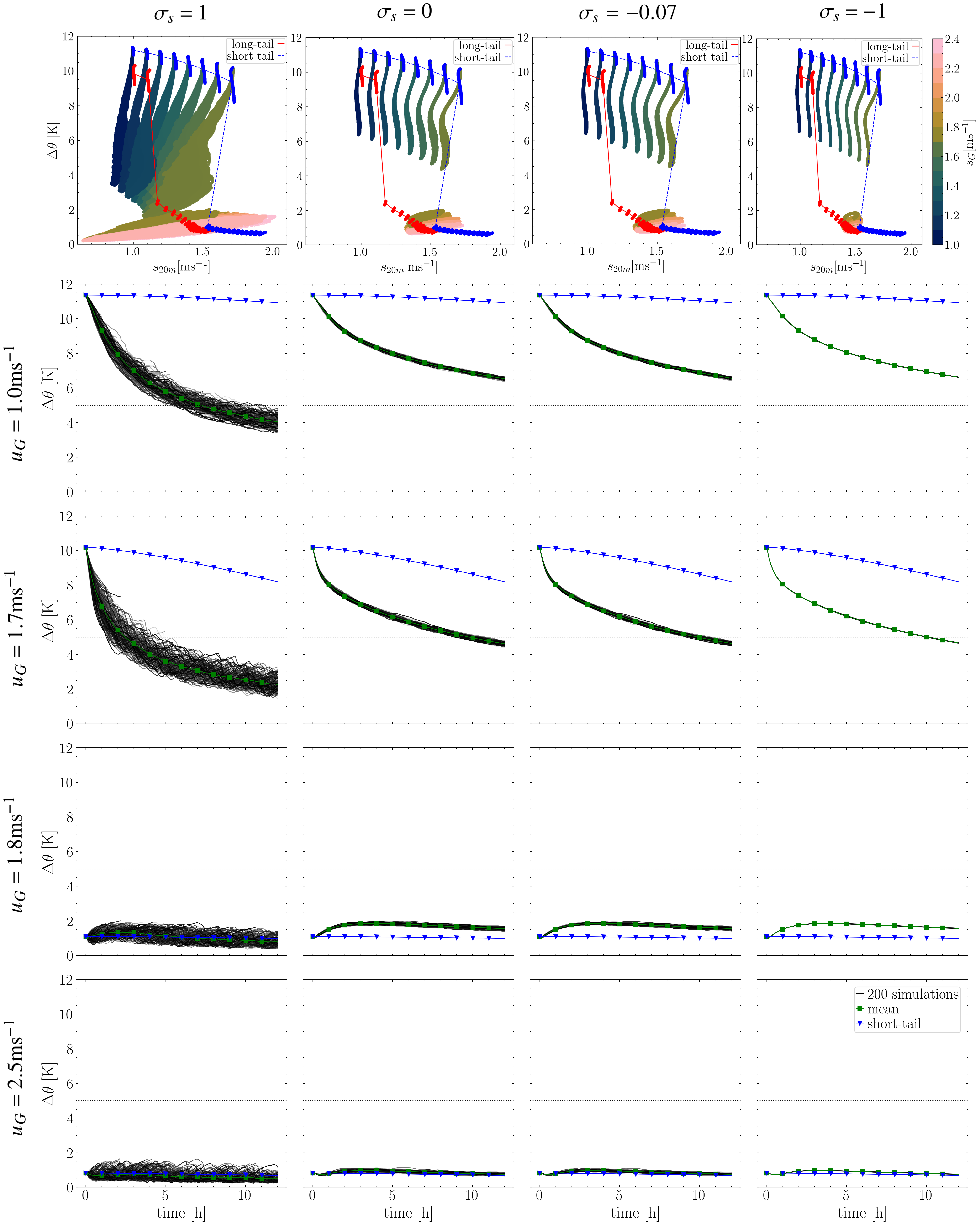}\\
  \caption{Results of the model (i.e., eq. \eqref{eq:pde_u} to \eqref{eq:pde_theta_g}) with a randomized stability function (eq. \eqref{eq:stoch_stab_func}) for the four different noise scenarios \ref{item:noise_scenario_1} - \ref{item:noise_scenario_4} (columns), and four different stability scenarios \ref{item:stability_scenario_sf1} - \ref{item:stability_scenario_sf4} (rows 2-4). Each combination of scenarios was run 200 times. The first row displays bifurcation plots for each noise scenario. The colors indicate the value of the geostrophic wind, and the blue dots and the blue line correspond to the results with a short-tail stability function and no added noise. Similarly, the red dots and the red line correspond to the results with a long-tail stability function and no noise. Rows 2-4 are plots of 200 simulations of $\Delta \theta$ over time, $t$, for all noise and stability scenarios. The green line is the mean for each scenario over all 200 simulations.}\label{fig:stab_func_results}
\end{figure}
\clearpage
For each noise and stability scenario, 200 simulations are run for 12 hours. The results are shown in figure \ref{fig:stab_func_results}. The columns in figure \ref{fig:stab_func_results} are the different noise scenarios. In the first row, the plots are bistability plots, similar to figure \ref{fig:bif_diagram}b, for all four noise scenarios. That means that the model with the stochastic stability equation is run for a range of geostrophic wind forcings. For each of these simulations, $\Delta \theta$ is plotted over the wind speed at 20~m. The geostrophic wind forcing corresponding to a simulation is indicated by color. As a comparison, $\Delta \theta$ from simulations with the deterministic short- and long-tail stability function from section \ref{sec:model}\ref{sec:model_description} (i.e., eqs. \ref{eq:short_tail_stab_func} and \ref{eq:long_tail_stab_func}) are plotted in blue and red.
Rows two to four show the result of all 200 simulations for each stability scenario and all noise levels. The black lines are the 200 simulations of $\Delta \theta$ plotted over time. The mean of these simulations is plotted in green, and as a comparison, $\Delta \theta$ from a simulation with the deterministic short-tail stability function (eq. \ref{eq:short_tail_stab_func}) is plotted in blue. The threshold of 5~K, which separates the very and weakly stable regime, is indicated with the horizontal dotted line.

The bifurcation diagram for the high-noise scenario shows that in all simulations with a geostrophic wind smaller than 1.8$~\mathrm{ms^{-1}}$, a very to weakly stable transition occurs. For none of the other noise scenarios can such strong transitions be observed. 

To get a clearer picture, the results for the four different stability and noise scenarios are studied in detail. Rows two to four show that in the high noise scenario (left column), $\Delta \theta$ displays a high variability, while for the scenario with the suppressed noise level (right column), the different simulations of $\Delta \theta$ are nearly equal to the mean over all simulations (green line). For the very stable scenario (\ref{item:stability_scenario_sf1}), only for the high noise scenario (\ref{item:noise_scenario_1}) transitions can be observed. All simulations from the other noise scenarios show a trend towards the weakly stable regime but the threshold is not crossed. For the scenario where the geostrophic wind is chosen such that it is the transition wind speed of the very stable regime (\ref{item:stability_scenario_sf2}), for all noise scenarios, the threshold is crossed, but apart from the high-noise scenario, the threshold is only barely crossed by the end of simulation time. Hence, they are not counted as full regime transitions. For none of the scenarios, there are weakly to very stable regime transitions.

The results show that with a stochastic stability equation, if the simulations are started in the very stable regime, there is a stronger trend towards the weakly stable regime compared to a model with a deterministic short-tail stability function. Hence, we conclude that a stochastic stability equation prevents the model from getting "stuck" in the very stable regime, thereby preventing the so-called "runaway cooling". However, weakly to very stable regime transitions do not occur when using this stability equation.

\section{Summary and Conclusion}
This study expands on prior research on the relevance of small-scale perturbations for SBL regime transitions. 
We used a TKE closure single-column model for the atmospheric boundary layer used in \cite{boyko_simulating_2023} to quantify the impact of perturbed tendencies and stochastic parameterizations. In particular, we studied the relevance of small-scale fluctuations of wind and temperature dynamics, as well as that of turbulence intermittency for regime transitions. We hypothesize that some limitations of using a TKE closure model regarding regime transitions can be counteracted by including perturbations. Specifically, temperature perturbations can induce buoyancy perturbations, which act as a source or sink in the TKE, depending on the sign of the perturbation. This could replace the source or sink that could otherwise arise from the inclusion of a prognostic equation for the sensible heat flux \citep{maroneze_new_2023}. The advantage of using a TKE closure model with perturbations instead of a model with a non-parameterized heat flux in NWPs is that it potentially reduces computational costs.

The fluctuations of the wind and temperature dynamics were included in the model by adding a 2D Gaussian function to either the evolution equation for the wind or the temperature dynamics. Different perturbation strengths and sizes were tested in a sensitivity study. The results for both perturbation types indicate that small-scale phenomena can drive regime transitions for all tested stability scenarios. For the perturbed temperature dynamics, persistent very to weakly stable regime transitions were observed if the geostrophic wind takes values within a wind speed transition region. Distinct but not permanent weakly to very stable transitions were observed for the perturbed temperature. In contrast, for the perturbed wind, the stability increased after the perturbation, but the regime threshold was barely, if at all, crossed. However, the results obtained with a negative wind perturbation must be treated with some reservation. The phenomena tested here are insufficient to cause persistent weakly to very stable regime transitions. Therefore, it is especially relevant to account for small-scale fluctuations in the wind and temperature dynamics in weather and climate models to achieve accurate regime transition statistics for very to weakly stable transitions. For the temperature dynamic perturbations, the time spread of the perturbation has a noticeable impact, while the height spread has less of an impact. For the perturbed wind dynamics, both dimensions are relevant.

In addition to the impact of the perturbations on regime transitions, we studied their effect on the dynamics of the SBL by studying the growth of the Ekman layer after a perturbation for different forcing scenarios. The results agreed with the previous findings that small-scale perturbations have a long-lasting impact for very to weakly stable transitions while for weakly to very stable they only have a transient impact. This occurs because top-down mixing events lead to a sudden growth of the Ekman layer that persists after it has occured.

Finally, to address model errors related to the Monin-Obukhov Similarity Theory, we investigated the influence of the stochastic stability equation, as initially defined by \cite{boyko_stochastic_2023}. The parameters of the stability correction were estimated based on observational data. Analogously, this methodology can be extended to Numerical Weather Prediction models or Earth system models to integrate localized turbulent bursts. Incorporating such transient events is of substantial scientific importance, given their potential to induce regime transitions within the stable boundary layer. Furthermore, utilizing a randomized stability function offers an alternative avenue to mitigate issues associated with the long-tail stability function by effectively reducing excessive mixing instead incorporating sporadic bursts of mixing. Simultaneously, a randomized stability equation could prevent the model from becoming trapped in a very stable state when employing a short-tail stability function. Our results show that the stochastic stability equation prevents the model from getting trapped in the very stable regime, thus preventing the so-called "runaway cooling". However, this stability equation does not facilitate weakly to very stable regime transitions. 

In conclusion, the here-tested perturbations and randomizations are a promising avenue to circumvent the need for excessive mixing through a long-tail stability function while also preventing runaway cooling in NWPs. The results suggest that small-scale fluctuations are most relevant for very to weakly stable regime transitions, which complements findings from \cite{abraham_climatological_2019} demonstrating that large-scale forcing is the main driver for weakly to very stable regime transitions, but less for very to weakly stable transitions.

This paper studies the effects of temperature and wind perturbations in isolation. Hence, an interesting research direction is to include combined wind and temperature perturbations. This study indicates a more pronounced response of the system to thermal perturbations, with both positive and negative temperature perturbations inducing a change in $\Delta \theta$ of approximately 13~K, in contrast to the $\approx$8~K change observed for positive wind perturbations. A discussion of negative wind perturbations is omitted due to the complications related to the unwanted change in the wind direction.
Based on these observations, we hypothesize that the change in the temperature has a dominating effect if the perturbations are combined. This implies that a warm micro front accompanied by an acceleration of wind would still lead to a decoupling of the layers, albeit at a diminished rate and intensity relative to that caused by a warm micro front combined with a wind deceleration. However, it should be noted that this effect depends on wind velocities that do not exceed a specific threshold. Similarly, a cold micro front coupled with low wind speeds is anticipated to couple the layers the most effectively.

Future work will entail making the perturbations of the wind and temperature dynamics stochastic and dependent on the stratification level and model resolution to make this approach applicable to NWPs.

%%%%%%%%%%%%%%%%%%%%%%%%%%%%%%%%%%%%%%%%%%%%%%%%%%%%%%%%%%%%%%%%%%%%%
% TABLES---INSERT NEAR IN-TEXT DISCUSSION
%%%%%%%%%%%%%%%%%%%%%%%%%%%%%%%%%%%%%%%%%%%%%%%%%%%%%%%%%%%%%%%%%%%%%
%%  Enter tables near where they are discussed within the document. 
%%  Please place tables before/after paragraphs, not within a paragraph.
%%
%
%\begin{table}[t]
%\caption{This is a sample table caption and table layout.  Enter as many tables as
%  necessary at the end of your manuscript. Table from Lorenz (1963).}\label{t1}
%\begin{center}
%\begin{tabular}{ccccrrcrc}
%\hline\hline
%$N$ & $X$ & $Y$ & $Z$\\
%\hline
% 0000 & 0000 & 0010 & 0000 \\
% 0005 & 0004 & 0012 & 0000 \\
% 0010 & 0009 & 0020 & 0000 \\
% 0015 & 0016 & 0036 & 0002 \\
% 0020 & 0030 & 0066 & 0007 \\
% 0025 & 0054 & 0115 & 0024 \\
%\hline
%\end{tabular}
%\end{center}
%\end{table}

%%%%%%%%%%%%%%%%%%%%%%%%%%%%%%%%%%%%%%%%%%%%%%%%%%%%%%%%%%%%%%%%%%%%%
% FIGURES---INSERT NEAR IN-TEXT DISCUSSION
%%%%%%%%%%%%%%%%%%%%%%%%%%%%%%%%%%%%%%%%%%%%%%%%%%%%%%%%%%%%%%%%%%%%%
%%  Enter figures near where they are discussed within the document.
%%  Please place figures before/after paragraphs, not within a paragraph.
% %
%
%\begin{figure}[t]
%  \noindent\includegraphics[width=19pc,angle=0]{figure01.pdf}\\
%  \caption{Enter the caption for your figure here.  Repeat as
%  necessary for each of your figures. Figure from \protect\cite{Knutti2008}.}\label{f1}
%\end{figure}

\clearpage
%%%%%%%%%%%%%%%%%%%%%%%%%%%%%%%%%%%%%%%%%%%%%%%%%%%%%%%%%%%%%%%%%%%%%
% ACKNOWLEDGMENTS
%%%%%%%%%%%%%%%%%%%%%%%%%%%%%%%%%%%%%%%%%%%%%%%%%%%%%%%%%%%%%%%%%%%%%
\acknowledgments
This project has received funding from the European Union's Horizon 2020 research and innovation program under the Marie Skłodowska-Curie grant agreement number 945371. The computations were performed on resources provided by Sigma2 - the National Infrastructure for High-Performance Computing and Data Storage in Norway. In addition, we thank the anonymous reviewers for their valuable and constructive feedback, which helped with
improving our paper.

%%%%%%%%%%%%%%%%%%%%%%%%%%%%%%%%%%%%%%%%%%%%%%%%%%%%%%%%%%%%%%%%%%%%%
% DATA AVAILABILITY STATEMENT
%%%%%%%%%%%%%%%%%%%%%%%%%%%%%%%%%%%%%%%%%%%%%%%%%%%%%%%%%%%%%%%%%%%%%
% 
%
\datastatement
The source code to reproduce all results can be found here: https://github.com/BoundaryLayerVercauteren/abl\_scm\_perturbation\_study.

%%%%%%%%%%%%%%%%%%%%%%%%%%%%%%%%%%%%%%%%%%%%%%%%%%%%%%%%%%%%%%%%%%%%%
% APPENDIXES
%%%%%%%%%%%%%%%%%%%%%%%%%%%%%%%%%%%%%%%%%%%%%%%%%%%%%%%%%%%%%%%%%%%%%
%
%% If only one appendix, use

%\appendix

%% If more than one appendix, use \appendix[<letter>], e.g.,

%\appendix[A] 

%% Appendix title is necessary! For appendix title:

%\appendixtitle{Title of Appendix}

%%% Appendix section numbering (note, skip \section and begin with \subsection)
%
% \subsection{First primary heading}

% \subsubsection{First secondary heading}

% \paragraph{First tertiary heading}

%%%%%%%%%%%%%%%%%%%%%%%%%%%%%%%%%%%%%%%%%%%%%%%%%%%%%%%%%%%%%%%%%%%%%
% REFERENCES
%%%%%%%%%%%%%%%%%%%%%%%%%%%%%%%%%%%%%%%%%%%%%%%%%%%%%%%%%%%%%%%%%%%%%
% Make your BibTeX bibliography by using these commands:
\bibliographystyle{ametsocV6}
\bibliography{references}

\end{document}